\documentstyle[a4,amstex]{article}

\newcommand{\cplx}{{\Bbb C}}

\newcommand{\m}{(m_1,m_2,\dots,m_M)}
\newcommand{\mi}{(m_{i_1},m_{i_2},\dots,m_{i_L})}
\renewcommand{\d}{(q-q^{-1})}
\newcommand{\glt}{U_q'(\hat{{\frak g}{\frak l}}_2)}

\newcommand{\Cz}{{\Bbb C}[z_1,\dots,z_N]}
\renewcommand{\S}{S^{\lambda}_N}
\newcommand{\mon}{z^{\l_{\s}}}
\newcommand{\moni}{z^{\l_{(i\,i+1)\s}}}
\newcommand{\zeti}{\zeta_i^{\l}(\s)}
\newcommand{\zetii}{\zeta_{i+1}^{\l}(\s)}
\newcommand{\qi}{q^{l_{\s_i}}}
\newcommand{\qii}{q^{l_{\s_{i+1}}}}
\newcommand{\setN}{\{1,2,\dots,N \}}
\newcommand{\setM}{\{1,2,\dots,M \}}
\newcommand{\seti}{\{i_1,i_2,\dots ,i_L\}}
\newcommand{\LM}{(0\leq L\leq M)}
\newcommand{\Sm}{S_N^{\m}}
\newcommand{\Smi}{S_{N,\mi}^{\m}}
\newcommand{\cV}{{\cal V}}
\newcommand{\cN}{{\cal N}}
\newcommand{\cZ}{{\cal Z}}
\newcommand{\D}{\Delta (u)}
\newcommand{\Ph}{\Phi_{\s}^{\l}}
\newcommand{\Phh}{\Phi_{(i,i+1)\s}^{\l}}
\newcommand{\ph}{\varphi_{\s}^{\l}}
\newcommand{\phh}{\varphi_{(i,i+1)\s}^{\l}}
\newcommand{\iNm}{i=1,\dots,N-1}
\newcommand{\Dou}{\Delta_0(u)}
\newcommand{\Du}{\Delta_1(u)}
\newcommand{\Dv}{\Delta_1(v)}

\newcommand{\Xu}{\Xi(u)}

\newcommand{\Xuo}{\Xi(u;\omega)}
\newcommand{\Xvo}{\Xi(v;\omega)}
\newcommand{\Tuo}{T_a^0(u;\omega)}
\newcommand{\Tvo}{T_a^0(v;\omega)}
\newcommand{\x}{\chi_{\s}}
\newcommand{\xx}{\chi_{(i,i+1)\s}}
\newcommand{\B}{{\cal B}}
\newcommand{\ti}{t_{i,i+1}}
\newcommand{\Mn}{{\frak M}_N}
\newcommand{\BU }{{\Bbb U}}
\newcommand{\BY}{{\Bbb Y}}

\renewcommand{\l}{\lambda}
\newcommand{\s}{\sigma}

\newcommand{\be}{\begin{enumerate}}
\newcommand{\ee}{\end{enumerate}}
\newcommand{\ba}{\begin{array}}
\newcommand{\ea}{\end{array}}
\newcommand{\beq}{\begin{equation}}
\newcommand{\eeq}{\end{equation}}
\newcommand{\bqa}{\begin{eqnarray}}
\newcommand{\eqa}{\end{eqnarray}}
\newcommand{\bqas}{\begin{eqnarray*}}
\newcommand{\eqas}{\end{eqnarray*}}
\newcommand{\halmos}{\rule{5pt}{5pt}}
\newcommand{\hecke}{H_N(q)}
\newcommand{\gi}{g_{i,i+1}}
\newcommand{\gij}{g_{i,j}}
\newcommand{\iN}{(i=1,\dots,N)}
\newcommand{\iNN}{(i=1,\dots,N-1)}

\renewcommand{\theequation}{\thesection.\arabic{equation}}
\numberwithin{equation}{section}
\newtheorem{df}{\bf Definition}
\newtheorem{prop}{\bf Proposition}

\newtheorem{lemma}{\bf Lemma}
\newtheorem{cor}{\bf Corollary}

\newenvironment{pf}{\noindent \it {\normalsize P}\normalsize{roof.}
\normalsize\hskip 5pt}{\hfill\halmos}

\begin{document}

\title{The trigonometric counterpart of the Haldane Shastry Model
}
\author{Denis Uglov \\ Research Institute for Mathematical Sciences\\
Kyoto University, Kyoto 606, Japan \\e-mail: duglov@@kurims.kyoto-u.ac.jp  }
\maketitle
\begin{abstract}
The hierarchy of Integrable Spin Chain Hamiltonians, which are trigonometric
analogs of the Haldane Shastry Model and the associated higher conserved
charges, is derived by a reduction from the trigonometric Dynamical Models of
Bernard-Gaudin-Haldane-Pasquier. The Spin Chain Hamiltonians have the property
of $\glt$-invariance. The spectrum of the Hamiltonians and the
$\glt$-representation content of their eigenspaces are found by a descent from
the Dynamical Models.
\end{abstract}

\section{Introduction}

\mbox{}

The Haldane Shastry Model [H,S] has been a subject of much attention in the
past years. This Model is a   version of the XXX Heisenberg Spin Chain where
spins interact with a potential inversely proportional to the squared distance
between the spins. Like the XXX Chain with nearest-neighbour interaction, the
Haldane Shastry Model is Integrable in the sense that its Hamiltonian is a
member of a family of mutually commuting, independent Integrals of Motion. The
Haldane Shastry Model, however, has been solved in much greater detail than the
nearest-neighbour XXX Model. The reason for this is a remarkably large algebra
of symmetries found in  the former Model. For the Haldane Shastry Spin Chain
with the spins taking values in the fundamental representation of ${\frak
g}{\frak l}_n$ the symmetry algebra is the ${\frak g}{\frak l}_n$-Yangian. This
infinite-dimensional algebra of symmetries facilitated  computation of explicit
expressions for the energy levels and the eigenvectors!
 [HHTBP,BPS,H2]. Some of the corre

lation functions have been found as well [HZ].

An important reason for the  attention attracted by the Haldane Shastry Model
is the fractional statistics exhibited by elementary excitations over the
antiferromagnetic ground state present in this Model [H2]. In the case of
${\frak g}{\frak l}_2$-Haldane Shastry Chain these excitations are spin-1/2
particles that are ``semions'' -- particles with statistics exactly half-way
between bosons and fermions in the sense of Haldane. Being exactly solvable the
Haldane Shastry Model provides a valuable laboratory for a study of physical
implications of the fractional statistics.

A remarkable connection exists between the Haldane Shastry Spin Chain in the
limit of infinite number of sites  and WZNW Conformal Field Theory at level 1
[BPS2,BLS]. This connection has led to a novel description of the space of
states in WZNW level-1 CFT, where the states are organized into irreducible
multiplets of Yangian symmetry algebra inhereted from the Haldane Shastry
Model. This, in turn, provided an explanation for the Fermionic Virasoro
character formulas for the level-1 integrable representations of $\hat{{\frak
s}{\frak l}}_2$ that were earlier derived by  [DKKMM].

In the paper [BGHP] it was realized that the Haldane Shastry Spin Chain is
related to a more general class of Integrable Models -- the so-called Dynamical
Models that describe quantum particles with spin moving along a circle. These
Dynamical Models can be thought of as generalzations of the Calogero-Sutherland
Model with inverse squared sine potential. The precise way in which the Haldane
Shastry Hamiltonian and associated higher conserved charges are obtained from
the hierarchy of  Dynamical Models was recently explained by Polychronakos in
[P], and by Talstra and Haldane in [TH]. In the last paper the authors have
shown how the hierarchy of Integrable Spin Models including the Haldane Shastry
Model appears in the static limit of the Dynamical Models in which  the
coordinates of the particles are ``frozen''  along the circle in an
equidistantly spaced lattice.

In the present paper we define trigonometric counterparts of the spin-1/2
Haldane Shastry Hamiltonian and the associated higher conserved charges. The
hierarchy of Integrable Spin Chain Hamiltonians that we obtain has an
infinite-dimensional symmetry algebra $\glt$ that takes over the role played by
the Yangian in the Haldane Shastry Model. We compute the eigenvalue spectrum of
the hierarchy and find that it has an additive, particle-like form. The space
of states is decomposed into eigenspaces of the operators that form the
hierarchy. Each of the eigenspaces is a highest-weight irreducible
representation of  $\glt$ parametrized by a sequence of integers --``magnon
quasimomenta'' or a ``motif'' in terminology of [HHTBP,BPS]. The eigenvalue
that corresponds to such eigenspace is a $q$-deformation of the eigenvalue of
the Haldane Shastry Hamiltonian parametrized by the same magnon quasimomenta.

The procedure that is used to derive the trigonometric hierarchy has been
inspired by the Talstra-Haldane approach -- we extract the Integrable Spin
Models from a static limit of the trigonometric  $\glt$-invariant Dynamical
Models that were defined in [BGHP]. The spectrum of the Spin Models is obtained
by a descent from the spectrum of these Dynamical Models.

\subsection{A survey of the method and results}
In this subsection we highlight the main steps of the procedure that is used to
define the trigonometric hierarchy of Integrable Spin Models and formulate the
results of this article. The details and proofs of the statements are contained
in the main body of the paper starting with sec.1.

We derive the hierachy  of trigonometric Spin Models by a two-step reduction
from the trigonometric Dynamical Models that were introduced in [BGHP]. First
of all we recall the definition of these trigonometric Dynamical Models.

\subsubsection{Trigonometric Dynamical Models of [BGHP]}

The trigonometric Dynamical Models are defined starting with two
representations of the finite-dimensional Hecke Algebra $\hecke$ (Cf. {\bf
1.1}). The first of these representations is defined in the ring of polynomials
in $N$ variables $\Cz$. The generators $g_{i,i+1} \quad \iNN$ of $\hecke$ in
this representation have the following form
\begin{equation}
 g_{i,i+1} :=\frac{q^{-1}z_i - qz_{i+1}}{z_i-z_{i+1}}(K_{i,i+1} - 1) + q
\qquad \iNN .
\end{equation}
Where $ K_{i,j} $ is the exchange operator for variables $ z_i , z_j $. \\
The second representation of $\hecke$ is defined in the $N$-fold tensor product
of two-dimensional vector spaces $H := V^{\otimes N},\; V:= \cplx^2 = \cplx\{
v^+ , v^-\}$. The Hecke generator $\ti \quad \iNN$ is a matrix acting in $H$
according  to the formula
\begin{equation}
\ti = I\otimes \dots \otimes I\otimes\underset{i,i+1}{t}\otimes I \otimes \dots
\otimes I  \quad \iNN,
\end{equation}
where $t$ is the matrix which acts in $ V\otimes V$:
\bqa
t v^+\otimes v^- & = & (q-q^{-1})v^+\otimes v^- + v^-\otimes v^+ \\
t v^-\otimes v^+ & = & v^+\otimes v^-  \\
t  v^\pm\otimes v^\pm & = & qv^\pm\otimes v^\pm.
\eqa
The two of these representations naturally extend to $ \Cz\otimes H$.

The Hecke representation generated by  $g_{i,i+1} \quad \iNN$ is enlarged to a
representation of the Affine Hecke Algebra $\widehat{\hecke}$ by adjoining
affine generators $ Y_i \quad \iN$ :
\bqa
Y_i & := & g_{i,i+1}^{-1}K_{i,i+1}\dots
g_{i,N}^{-1}K_{i,N}p^{D_i}K_{1,i}g_{1,i}\dots K_{i-1,i}g_{i-1,i}.
\eqa
Where $ p $ is a $c$-number , $ D_i := z_i \frac{\partial}{\partial z_i} $ and
\begin{equation}
 g_{i,j} :=\frac{q^{-1}z_i - qz_j}{z_i-z_j}(K_{i,j} - 1) + q  \qquad (i,j =
1,\dots,N-1) .
\end{equation}

After introducing the objects just described the authors of [BGHP] define the
hierarchy of Dynamical Models. The operators $ \Delta^{(n)} \quad (n=1,\dots,
N)$ that constitute the hierarchy are coefficients of the polynomial $\D =
\sum_{n=0}^N u^n  \Delta^{(n)}$ which generates elementary symmetric functions
of $ Y_1,Y_2,\dots,Y_N $ :
\begin{equation}
\D : = \prod_{i=1}^N ( 1 + u Y_i ) .
\end{equation}
Due to the mutual commutativity of the operators $ Y_1,Y_2,\dots,Y_N $ the
hierachy of Dynamical Models is integrable:
\begin{equation}
 [ \Delta(u) , \Delta(v) ] = 0 .
\end{equation}

\mbox{}

In the space $ \Cz\otimes H $ one defines a representation of the algebra $
\glt $ (Cf. {\bf 1.2}). In this representation the generators of $ \glt $ are
obtained by expanding in the parameter $u$ the monodromy matrix $ T_a(u) $
which is defined in a standard way as a product of elementary $L$-operators
[BGHP]:
\begin{equation}
T_a(u) : = L_{a1}(uY_1)L_{a2}(uY_2)\dots L_{aN}(uY_N).
\end{equation}
Where the elementary $L$-operator  $L_{ai}(uY_i)\quad \iN$ acts in the tensor
product of an auxiliary copy of the two-dimensional vector space denoted by
$V_a$ and the space $ \Cz\otimes H$:
\bqa
L_{ai}(uY_i) & := & \frac{uY_it_{a,i} - t_{a,i}^{-1}}{uY_i - 1}P_{a,i} \quad
\iN.
\eqa
Here $P$ is the permutation operator in $ V\otimes V$ and $ t_{a,i} \, , \,
P_{a,i} $ are the usual extensions of $ t \, , \, P$ as operators in $
V_a\otimes H$.
The fact that $ T_a(u) $ defines a representation of $ \glt $ follows from the
$RTT = TTR$ relation which involves the trigonometric $R$-matrix:
\begin{align}
\bar{R}_{ab}(u/v)T_a(u)T_b(v) & =  T_b(v)T_a(u))\bar{R}_{ab}(u/v), \\
\bar{R}(z) & :=   \frac{ zt - t^{-1} }{qz - q^{-1}}P.
\end{align}

The hierarchy of Dynamical Models defined by $\D$ is $\glt$-invariant, that is
\begin{equation}
[ \D , T_a(v) ] = 0 .
\end{equation}
This again follows from the mutual commutativity of the operators
$Y_1,Y_2,\dots,Y_N.$

Both $\D$ and $ T_a(u) $ act in the ``bosonic'' subspace of $ \Cz\otimes H $ as
explained in [BGHP]. This subspace which we denote by $\B$ is defined by the
requirement of the Hecke-invariance:
\begin{equation}
\B := \{ b \in \Cz\otimes H |  (\gi - \ti)b = 0 \quad \iNN \}.
\end{equation}

In any operator $O$ acting in $\B$ one can eliminate the coordinate exchange
operators $K_{i,j}$ by carrying them one-by-one to the right of any expression
in $O$ and replacing a $ K_{i,j} $ standing on the right of an expression in
accordance with the rule: $ \gi \rightarrow \ti \quad \iNN $. This leads to a
uniquely defined operator $\widehat{O}$ which does not contain coordinate
permutations, such that
\begin{equation}
O\B = \widehat{O}\B .
\end{equation}
Where the notation means that the equality holds for any vector in $\B$.\\
With this definition the eq. (0.0.9,.12,.14) lead to
\begin{align}
[ \widehat{\Delta(u)} , \widehat{\Delta(v)}]\B  & = 0 , \\
[ \widehat{\D} , \widehat{T_a(v)} ]\B & = 0 , \\
(\bar{R}_{ab}(u/v)\widehat{T_a(u)}\widehat{T_b(v)} & -
\widehat{T_b(v)}\widehat{T_a(u)}\bar{R}_{ab}(u/v))\B = 0 .
\end{align}
The set of relations (0.0.17-.19) constitutes the  result of [BGHP] concerned
witn the trigonometric Dynamical Models.

\subsubsection{The hierarchy of trigonometric Dynamical Models at $p=1$}

The first step of reduction from the Dynamical to the Spin Models consists in
taking the static limit $ p\rightarrow 1$ in the construction described in the
previous subsection. Following the idea of [TH] we expand the generating
function for the commuting charges of the hierarchy around the point $p=1$ up
to the linear term in $p-1$:
\begin{equation}
\D = \Dou + (p-1) \Du + O ( (p-1)^2).
\end{equation}
In the symmetry generator $ T_a(u) $ we retain the leading term only:
\begin{equation}
T_a(u) = T_a^0(u) + O(p-1).
\end{equation}
The operators $ \Dou $ and $ \Du $ turn out to have a very special form.
First of all $ \Dou $ is a {\em constant }:
\begin{equation}
\Dou = \prod_{i=1}^N(1 + uq^{2i - N -1}).
\end{equation}
({\bf Remark} In the paper [TH] which deals with the rational case, the leading
term of the generating function
\begin{align*}
\Delta_{rational}(u) & := \prod_{i=1}^N(u - d_i) ,  \\
(d_i & := hD_i +
\sum_{j>i}\frac{z_i}{z_i-z_j}K_{i,j}-\sum_{j<i}\frac{z_j}{z_j-z_j}K_{i,j} \quad
\iN )
\end{align*}
in the static limit $ h\rightarrow 0 $ is not a constant but a function of
$z_1,\dots,z_N$. This is due to a choice of the Dunkl operators $d_i$ (above)
which do not act in the space of polynomials. The rational limit of the affine
Hecke generators $ Y_i $ which we use gives the gauge transformed Dunkl
operators acting in the space of polynomials ( Cf. [BGHP]).)

Since $\Dou$ is a constant, the  first-order differential operator $ \Du $
takes over the role of the generating function for Integrals of Motion:
\begin{align}
[\Du , \Dv ]&  = 0, \\
[\Du , T^0_a(v)] & = 0 .
\end{align}
Secondly, we find that $\Du$ has the following structure:
\begin{equation}
\Du = \sum_{i=1}^N \theta(u;z)_iD_i  + \Xi(u;z) .
\end{equation}
Where $ \Xi(u;z) $ is a function of the operators $ z_1,\dots,z_N $ and $
K_{i,j} \quad (i,j = 1,\dots,N ) $  only. On the other hand the coefficients $
\theta(u;z)_i $ do not depend on the operators of coordinate permutation and
are functions of the coordinates $ z_1,\dots,z_N  \quad \iN $ only. This kind
of separation of the differentials $ D_i $ and the operators of coordinate
permutation was first observed by [TH] in the rational case.

For the differential part  $ {\cal D}(u) := \sum_{i=1}^N \theta(u;z)_iD_i $ of
$\Du $ we obtain an explicit expression in terms of the generating function $
D(u;p,t)$ for the Macdonald operators [M],[JKKMP] (Cf. {\bf 2.1}):
\begin{equation}
{\cal D}(u) = D_1(q^{N-1}u;q^{-2}) .
\end{equation}
Where  $ D_1(u;t)$ is the linear term in the expansion of the generating
function  $ D(u;p,t)$
around the point $ p=1$ :
\begin{equation}
D(u;p,t) = \Delta_0(u;t) + (p-1)D_1(u;t) + O((p-1)^2).
\end{equation}

The zero-order part $\Xi(u;z)$ of the differential operator $ \Du $ is the
object which we use to define the hierarchy of Integrable Spin Models.

First of all we observe that $ \Du $ lies in the centre of the Affine Hecke
Algebra generated by
$ \gi \quad \iNN $ and $ y_i := Y_i|_{p=1} \quad \iN $. Therefore $ \Du $  acts
in the bosonic subspace $\B$ defined in the previous subsection, and in this
subspace we have analogs of the relations (0.0.17,-.19):
\begin{align}
[ \widehat{\Du} , \widehat{\Dv}]\B  & = 0 , \\
[ \widehat{\Du} , \widehat{T^0_a(v)} ]\B & = 0 , \\
(\bar{R}_{ab}(u/v)\widehat{T^0_a(u)}\widehat{T^0_b(v)} & -
\widehat{T^0_b(v)}\widehat{T^0_a(u)}\bar{R}_{ab}(u/v))\B = 0 .
\end{align}
The operator $\widehat{\Du}$ is a sum of two parts: the $ {\cal D}(u) $ which
is a first order differential operator and $ \widehat{\Xi(u;z)}$ which is a
matrix acting on $H$ whose entries are rational functions of the coordinates $
z_1,\dots,z_N .$  At the second step of the reduction from the Dynamical Models
to the Spin Models we shall eliminate the differential part of $\widehat{\Du}$
by restricting the coordinates to special values in a way which leaves the
Integrability and the $ \glt $-invariance intact.

\subsubsection{Definition of the hierarchy of trigonometric Spin Models}

The way to ``freeze'' the coordinates while keeping the spins as dynamical
variables goes trough the use of the {\em evaluation map } $Ev(v): \Cz\otimes H
\mapsto H $. This map is parametrized by complex numbers $ v_1,\dots, v_N $ and
works by taking values of functions of $z_1,\dots,z_N $ at the point $ z_1 =
v_1 ,\dots,z_N = v_N $. We use this map at the special point $z = \omega  : z_1
= \omega^1,\dots,z_N = \omega^N $ where $ \omega = e^{2\pi i /N}.$

In order to explain the relevance of this point we first of all observe, that
at this point the coefficients of the differential operator $ {\cal D}(u) $ are
all equal one to another:
\begin{equation}
 \theta(u;\omega)_i = \theta(u)  \qquad \iN .
\end{equation}
Where the constant $\theta(u)$ is given by eq. (2.2.26) in the main text. Next,
we observe, that in the expansion of $\widehat{\Du}$ in $u$: $ \widehat{\Du}:=
\sum_{n=1}^N u^n \Hat{\Delta}_1^{(n)} $ the term $\Hat{\Delta_1^{(N)}}$ is the
scale operator: $\Hat{\Delta}_1^{(N)} = D_1 + D_2 + \dots + D_N$. We can modify
the generating function $ \widehat{\Du}$ by subtracting from it the product of
the constant $\theta(u)$ and the operator $\Hat{\Delta}_1^{(N)}$. The equations
(0.0.28 -.30) clearly still hold for this modified generating function $
\widehat{\Du} - \theta(u)\Hat{\Delta}_1^{(N)} $. Moreover due to (0.0.31) for
any vector $ b \in \B $ we have:
\begin{equation}
Ev(\omega) (\widehat{\Du} - \theta(u)\Hat{\Delta}_1^{(N)})b = Ev(\omega)
\widehat{\Xi(u;z)}b .
\end{equation}
The map $Ev(\omega)$ naturally pulls through the operators $\widehat{\Xi(u;z)}$
and $ \widehat{T_a^0(u)} $ since these operators are matrices acting in $H$,
and the entries of these matrices are rational functions of $z \equiv (
z_1,\dots,z_N )$ non-singular at the point $z = \omega$. By pulling
$Ev(\omega)$ through $\widehat{\Xi(u;z)}$ and $ \widehat{T_a^0(u)}$ we define
operators $ \Xuo $ and $ T_a^0(u;\omega) $ acting in the image of $ Ev(\omega)
$ in $H$ by :
\begin{align}
Ev(\omega)\widehat{\Xi(u;z)}\B & = \Xuo Ev(\omega)\B , \\
Ev(\omega)\widehat{T_a^0(u)}\B & = T_a^0(u;\omega) Ev(\omega)\B .
\end{align}

Taking (0.0.31) and (0.0.33,.34) into account we apply the evaluation map to
the relations (0.0.28 - .30) and get:
\begin{align}
[ \Xuo , \Xvo]H_{\B}(\omega)  & = 0 , \\
[ \Xuo , T^0_a(v;\omega) ]H_{\B}(\omega) & = 0 , \\
(\bar{R}_{ab}(u/v)T^0_a(u;\omega)T^0_b(v;\omega) & -
T^0_b(v;\omega)T^0_a(u;\omega)\bar{R}_{ab}(u/v))H_{\B}(\omega) = 0 .
\end{align}
Where $ H_{\B}(\omega):= Ev(\omega)\B \subset H .$

Next, we  prove, that $H_{\B}(\omega) = H $ .Therefore the relations (0.0.35
-.37) express Integrability and $ \glt$-invariance of the hierarchy of Spin
Chain Models. The  Hamiltonians of these Models act in $H$ and are obtained by
expanding the generating function $ \Xuo $ in the parameter $u$:
\begin{equation}
\Xuo = \sum_{n=1}^{N-1} u^n \Xi^{(n)}(\omega).
\end{equation}

While the generating function $ \Xuo $ is completely defined by the relation
(0.0.25), the computation of an explicit expression is still quite a difficult
task. We have computed the explicit expression only for the first member of the
hierarchy -- the operator $ \Xi^{(1)}(\omega)$. To give this expression we
introduce several notations. For $ i\neq j \in \{1,\dots,N-1\}$   define the
rational functions:
\begin{equation}
a_{i,j} := \frac{q^{-1}z_i - qz_j}{z_i-z_j}\; , \quad   b_{i,j}  :=
\frac{(q-q^{-1})z_i}{z_i-z_j}.
\end{equation}
Define also the matrices:
\begin{multline*}
Y_{i,i+1}(w)  :=  \frac{ w\ti - \ti^{-1} }{qw - q^{-1}}, \\
M^{(j,i)}(x,y)  := \left( Y_{i+1,i+2}(y/z_{i+1})\dots
Y_{j-1,j}(y/z_{j-1})\right)^{-1}Y_{i,i+1}(x/y) \times \\ \times \left(
Y_{i+1,i+2}(x/z_{i+1})\dots Y_{j-1,j}(x/z_{j-1})\right) \quad (j > i).
\end{multline*}
With these notations we have:
\begin{multline}
\Xi^{(1)}(\omega) = \\
\sum_{M=2}^N \frac{(-1)^M}{\d} \sum_{N \geq i_M > \dots > i_1 \geq 1} {\cal
H}(\omega)_{i_M,i_{M-1},\dots,i_1} R(\omega)_{i_2,i_1}R(\omega)_{i_3,i_2 \dots
}R(\omega)_{i_M,i_{M-1}} + cI ; \\
{\cal H}(z)_{i_M,i_{M-1},\dots,i_1}:= \left( \prod_{i_1 < f <
i_2}a_{i_1,f}\right)\left( \prod_{i_2 < f < i_3}a_{i_2,f} \right)\dots
\left(\prod_{i_M < f < N + i_1}a_{i_M,f (\mod N)}\right) \times \\
\times b_{i_M,i_{M-1}} b_{i_{M-1},i_{M-2}}\dots  b_{i_2,i_1} b_{i_1,i_M}, \\
R(z)_{j,i} :=  M^{(j,i)}(z_{i_M},z_i) \qquad (N \geq j > i \geq 1).
\end{multline}
Where $c$ is unimportant constant.
In the limit $q \rightarrow 1 $ we recover the Haldane Shastry Hamiltonian:
\begin{equation}
 \lim_{q\rightarrow 1}\frac{\Xi^{(1)}(\omega)}{\d} = H_{HS} := -\sum_{N \geq j
> i \geq 1} \frac{\omega^i \omega^j}{ (\omega^i - \omega^j)^2 } (P_{i,j} - 1).
\end{equation}
While the Hamiltonian $\Xi^{(1)}(\omega)$ looks rather intimidating, its
spectrum, as well as the spectrum of the whole hierarchy generated by $\Xuo$
has a remarkably simple additive form. We describe this spectrum in the next
subsection.

\subsubsection{Eigenvalue spectrum of the hierarchy of Spin Hamiltonians}

As in the case of the Haldane Shastry Model the common eigenspaces of the
operators $\Xi^{(n)}(\omega)\quad (n=1,\dots,N-1)$ are in one-to-one
correspondence with ${\frak sl}_2$ motifs (Cf. [HHTBP] or [BPS]). A  sequence
of integers $\m $  is called an ${\frak sl}_2$ motif iff :
\begin{align}
 & 1 \leq m_1 < m_2 < \dots < m_M \leq N-1 \, ; \tag{\theequation a.}  \\
 &  m_{i+1} - m_i \geq 2  \quad ( i = 1,\dots , M-1 ) \tag{\theequation b.}.
\end{align}
For a fixed $N$ we denote the set of all $\frak{sl}_2$ motifs including the
empty one by $\Mn$. For the eigenspace of $\Xuo$ which correspons to a motif
$\m $ we use the notation $ H_{\B}^{\m}(\omega)$. We have:
\begin{equation}
\Xuo H_{\B}^{\m}(\omega) = \left( \sum_{i=1}^M \xi^{(m_i)}(u)
\right)H_{\B}^{\m}(\omega).\end{equation}
Where the sum is understood to be zero for $ M=0$. \\
The elementary eigenvalue $\xi^{(m)}(u) \quad  (m \in \{1,\dots,N-1\})$ is
\begin{equation}
\xi^{(m)}(u) = u\prod_{k=1}^N(1 + u q^{2k-N-1})\left\{\sum_{i=1}^m \frac{q^{2i
- N-1}}{1+ uq^{2i - N-1}} -\frac{m}{N} \sum_{i=1}^N \frac{q^{2i - N-1}}{1+
uq^{2i - N-1}} \right\}.
\end{equation}
In particular the elementary eigenvalue of the Hamiltonian $\Xi^{(1)}(\omega)$
is given by
\begin{equation}
\xi^{(m),(1)} = \frac{q^{-N}}{N}( Nq^m[m]_q  - mq^N[N]_q ).
\end{equation}
Where we used the usual notation $ [x]_q \equiv \frac{q^x - q^{-x}}{\d}$. In
the limit $ q\rightarrow 1 $ we recover the elementary eigenvalue of the
Haldane Shastry Model [HHTBP,BPS]:
\begin{equation}
\lim_{q\rightarrow 1}\frac{\xi^{(m),(1)}}{\d} = m(m - N).
\end{equation}

The space of states of the Spin Models is represented as  a  direct sum of the
eigenspaces  $ H^{\m}(\omega)$:
\begin{equation}
 H = \bigoplus_{\m \in \Mn }H_{\B}^{\m}(\omega).
\end{equation}

\subsubsection{Structure of the common eigenspaces of the hierarchy of Spin
Hamiltonians.}

Each of the eigenspaces $ H_{\B}^{\m}(\omega)$ is an irreducible highest-weight
representation of $\glt$. The Drinfeld polynomial [CP] $Q^{\m}(u)$ of $
H_{\B}^{\m}(\omega)$ is
\begin{equation}
Q^{\m}(u) = \prod\begin{Sb}1\leq  k \leq N \\  k \not\in \{m_i,m_i+1\} \end{Sb}
(1 - q^{-2k+N+1}u).
\end{equation}
(Cf. {\bf 1.2 } for our conventions about Drinfeld polynomials ).

To describe an eigenspace $ H_{\B}^{\m}(\omega) \quad \m \in \Mn $ in a more
explicit way we give some facts (Cf. {\bf 4.1}) about eigenvectors of the
operator $ \Du $ which defines the hierarchy of Dynamical Models at $p=1$.

The linear space of polynomials $\Cz$ is represented as a direct sum of
eigenspaces of the operator $\Du $. There is a one-to-one correspondence
between these eigenspaces and partitions with $ N$ parts. In an eigenspace
which corresponds to a partition $ \l := ( \l_1 \geq \l_2 \geq \dots \l_N \geq
0 ) $ there exists a basis $\{ \ph(z) \}_{\s \in \S}$. The polynomials  forming
 this basis are indexed by elements from symmetric group $S_N$. There is a
one-to-one correspondence between the elements of the basis $\{ \ph(z) \}_{\s
\in \S}$ and elements of the set $ \S \quad (\S \subset S_N )$ defined in {\bf
1.4}. A polynomial $  \ph(z) \quad (\s \in \S) $ is completely specified by the
three conditions:\begin{align}
\Du \ph(z) & = \left(\prod_{k=1}^N(1 + u q^{2k-N-1})\left\{\sum_{i=1}^N
\frac{uq^{2i-N-1}}{1 + uq^{2i-N-1}}\l_i \right\} \right) \ph(z) ,
\tag{\theequation a.} \\
y_i \ph(z) & = q^{2\s_i - N -1} \ph(z) \quad \iN , \tag{\theequation b.} \\
\ph(z) & = z_1^{\l_{\s_1}}z_2^{\l_{\s_2}} \dots z_N^{\l_{\s_N}} + \text{smaller
monomials}.\tag{\theequation c.}
\end{align}
We remind, that $y_i := Y_i|_{p=1} \quad \iN $. \\
The ``smaller monomials `` means a linear combination of monomials that are
smaller than the monomial $z_1^{\l_{\s_1}}z_2^{\l_{\s_2}} \dots
z_N^{\l_{\s_N}}$ in the ordering described in {\bf 1.4(2)} (Cf. [BGHP]).

With any motif $ \m \in \Mn $ associate the partition
\begin{multline}
 (\underset{1}{M},\dots,\underset{m_1}{M},\underset{m_1+1}{M-1},\dots
,\underset{m_2}{M-1},\underset{m_2+1}{M-2},  \ldots \\ \ldots,
\underset{m_M}{1},\underset{m_M+1}{0},\dots ,\underset{N}{0} )
\end{multline}
We use  the same notation $ \m $ for a motif and the associated partition.

For an element $\s $ of the symmetric group $S_N$ define
\begin{align}
\{\s_1 , \s_2 , \dots ,\s_N \} & := \s.\setN \, ;  \tag{\theequation a.} \\
 i & := \s_{p^{\s}_i} \qquad \iN . \tag{\theequation b.}
\end{align}
For any $\m \in \Mn $ define the subset $S^{\m}_{N,\m}$ of $S_N$ (Cf. {\bf
4.3(2)}):
\begin{multline}
 S^{\m}_{N,\m} := \\ \left\{ \s \in S_N \quad \begin{array}{| c c c } p^{\s}_i
<  p^{\s}_{i+1} &  ( m_k  < i < m_{k+1}) & \text{for all}\; k \in
\{0,1,\dots,M\} \\
 p^{\s}_{m_k} >  p^{\s}_{m_k+1} &  & \text{for all} \; k  \in \{1,\dots,M-1\}
\end{array} \right\}.
\end{multline}
Where we adopt the convention: $ m_0 := 0 \, , m_{M+1} := N+1 .$

Next, for any motif $ \m \in \Mn $ we define  the subspace $W^{\m}$ of the
space of states $H$ (Cf. {\bf 6.1(2) }) as follows:
\begin{multline}
W^{\m}:= \\  S_q(\underset{1}{V}\otimes\dots\otimes\underset{m_1-1}{V})\otimes
A_q(\underset{m_1}{V}\otimes\underset{m_1+1}{V})\otimes
S_q(\underset{m_1+2}{V}\otimes\dots\otimes\underset{m_2-1}{V})\otimes
A_q(\underset{m_2}{V}\otimes\underset{m_2+1}{V})\otimes \ldots \\ \ldots
\otimes
S_q(\underset{m_{M-1}+2}{V}\otimes\dots\otimes\underset{m_M-1}{V})\otimes
A_q(\underset{m_M}{V}\otimes\underset{m_M+1}{V})\otimes
S_q(\underset{m_M+2}{V}\otimes\dots\otimes\underset{N}{V}) \\
W^{\m} \subset H:= \underset{1}{V}\otimes\underset{2}{V}\otimes \ldots
\underset{N}{V}.
\end{multline}
Where $ S_q $ and $ A_q$ mean $q$-symmetrization and $q$-antisymmetrization as
defined in (1.1.14,.15,5.5.17).
The space $W^{\m} \quad \m \in \Mn$ is an irreducible highest weight
$\glt$-module. The Drinfeld polynomial of this module is given by (0.0.47). The
  $\glt$-action on $W^{\m}$ is given by
\begin{equation}
{\cal L}(u;\{q^{2\s[0]_i - N -1}\}):= L_{a1}(uq^{2\s[0]_1 - N
-1})L_{a2}(uq^{2\s[0]_2 - N -1})\dots L_{aN}(uq^{2\s[0]_N - N -1}).
\end{equation}
Where for a fixed $\m$ we have introduced the notation
\begin{equation}
\s[0] := (m_1,m_1+1)(m_2,m_2+1)\dots(m_M,m_M+1) \in S_{N,\m}^{\m} \subset S_N.
\end{equation}

The eigenspace $ H_{\B}^{\m}(\omega) \quad (\m \in \Mn) $ is isomorphic to $
W^{\m}$ as $\glt$-module. This isomorphism is given by an invertible
intertwiner $ \check{\BU }^{\m}(\omega)$
\begin{equation}
H_{\B}^{\m}(\omega) = \check{\BU }^{\m}(\omega)W^{\m}.
\end{equation}
The intertwiner $\check{\BU }^{\m}(\omega)$ is defined by the expression
\begin{equation}
\check{\BU}^{\m}(\omega) := (-(q^2 + 1))^M\sum_{\s \in
S_{N,\m}^{\m}}\varphi_{\s}^{\m}(\omega)\BY(\s) .
\end{equation}
In this definition
\begin{equation}
\varphi_{\s}^{\m}(\omega) = \ph(z)|_{z_1=\omega^1,\dots,z_N=\omega^N}.
\end{equation}
Where the partition $ \l$ is the one specified by $ \m $ in accordance with
(0.0.48).
The matrix $\BY(\s) \quad ( \s \in S_{N,\m}^{\m})$ is an intertwiner which is
defined by the recursion relations (Cf. {\bf 5.2(4)} ):
\begin{gather*}
\BY(\s[0]):= \text{Id} \; ,  \\
\BY((i,i+1)\s) :=  \begin{cases} Y^+_{i,i+1}(q^{2\s_i-2\s_{i+1}})\BY(\s) &
\text{ if  $ \s_i - \s_{i+1} \geq 2 $}, \\
Y^-_{i,i+1}(q^{2\s_i-2\s_{i+1}})\BY(\s) & \text{ if  $  \s_i - \s_{i+1} \leq -2
$ }.
\end{cases}
\end{gather*}
Where the matrices $ Y^{\pm}_{i,i+1}(w) $ are
\begin{align}
Y^{\pm}_{i,i+1}(w) & := \varrho^{\pm}(w)\frac{w\ti - \ti^{-1}}{q^{-1}w - q } \,
,  \\
\varrho^+(w) & := \frac{w -1}{q^2 w -1} \;, \qquad \varrho^-(w)  := \frac{w -
q^2}{ w -1}. \notag
\end{align}
Notice that  $\BY(\s) \quad ( \s \in S_{N,\m}^{\m})$ is an invertible
intertwiner.

In general we cannot claim to know the explicit expression for the intertwiner
$\check{\BU}^{\m}(\omega)$ since we have not found the eigenvectors $
\varphi_{\s}^{\m}(z) $ explicitely. One exception is the case $q=0$ when
$\check{\BU}^{\m}(\omega)$ becomes very simple. In this case we have
\begin{equation}
\check{\BU}^{\m,q=0}(\omega)|_{W^{\m,q=0}} =
\omega^{\frac{1}{2}\sum_{i=1}^Mm_i(m_i + 1)}{\mathrm {Id}}.
\end{equation}
Therefore at $q=0$ the eigenspace $H_{\B}^{\m}(\omega)$ is the linear span of
the following vectors in $H$:
\begin{multline*}
\left|\left\{\begin{array}{c c c c c
}\overset{1}{+}&\overset{}{+}&\overset{}{+}&\dots &\overset{m_1-1}{+} \\
                    -& +& +& \dots & + \\
                    -& -& +& \dots & +  \\
                     \vdots  & \vdots  & \vdots & & \vdots    \\
                    - & - & - &  \dots & - \end{array} \right\} \right.
\overset{m_1}{+}\overset{m_1+1}{-}
\left\{\begin{array}{c c c c c
}\overset{m_1+2}{+}&\overset{}{+}&\overset{}{+}&\dots &\overset{m_2-1}{+} \\
                    -& +& +& \dots & + \\
                    -& -& +& \dots & +  \\
                     \vdots  & \vdots  & \vdots & & \vdots    \\
                    - & - & - &  \dots & - \end{array} \right\}
\overset{m_2}{+}\overset{m_2+1}{-}  \ldots \\ \ldots
\overset{m_M}{+}\overset{m_M+1}{-}
\left.\left\{\begin{array}{c c c c c
}\overset{m_M+2}{+}&\overset{}{+}&\overset{}{+}&\dots &\overset{N}{+} \\
                    -& +& +& \dots & + \\
                    -& -& +& \dots & +  \\
                     \vdots  & \vdots  & \vdots & & \vdots    \\
                    - & - & - &  \dots & - \end{array} \right\} \right\rangle
\end{multline*}
Where we use the notation:
\begin{equation}
|\epsilon_1 \epsilon_2 \dots \epsilon_N \rangle := v^{\epsilon_1}\otimes
v^{\epsilon_2}\otimes \dots \otimes v^{\epsilon_N} \quad ( \epsilon_i = \pm )
{}.
\end{equation}
For example when $N$ is even, $H_{\B}^{(1,3,\dots,N-1)}(\omega)$ is
one-dimensional and at $q=0$ it is spanned by the vector (antiferromagnetic
ground state):
\begin{equation*}
|+ - + - \dots + - \rangle
\end{equation*}

\mbox{}

\noindent In the rest of the paper we give a detailed exposition of the matters
which were briefly recounted in this introduction. In the sec.1 we gather
predominantly known facts about the trigonometric Dynamical Models of [BGHP]
and explain the conventions about the algebra $ \glt $ that we use. In the sec.
2 we discuss the hierarchy of Dynamical Models in the static limit $ p=1$. In
the sec. 3 we define the hierarchy of the Spin Models. The sec. 4. is concerned
with properties of the eigenvectors of the hierarchy of Dynamical Models in the
limit $p = 1$. In the sec. 5 we construct the Hecke-invariant ``bosonic''
eigenspaces for the Dynamical Models with spin at $p=1$. The eigenvalue
spectrum of the Spin Models and the $\glt $-representation content of their
eigenspaces are derived in sec. 6.

\mbox{}

\noindent \begin{Large}{\bf Acknowledgments}\end{Large}\\  I am most grateful
to Drs. R. Kedem and R. Weston and to Professors M.Jimbo, M.Kashiwara and
T.Miwa for numerous discussions and support.

\section{The $\glt$-invariant Dynamical Models}
In this section we summarize largely known facts about the trigonometric
Dynamical Models defined by [BGHP]. We also recount several facts about the
algebra $\glt$ and explain our notations.

\subsection{The representations of the Affine Hecke Algebra}

\subsubsection{The representation of $ \widehat{\hecke} $ in the space of
polynomials}

Following [BGHP] define the operators $ g_{i,j} \in End( \Cz)  \quad (i,j
=1,\dots, N) $:

$$  g_{i,j} := a_{i,j}K_{i,j}+ b_{i,j} $$

where $$ a_{i,j} = \frac{q^{-1}z_i - qz_j}{z_i-z_j}\;, \qquad b_{i,j} =
\frac{(q-q^{-1})z_i}{z_i-z_j} = q - a_{i,j}. $$ $ K_{i,j} $ is the interchange
operator for variables $ z_i , z_j $.

For $ p \in \cplx $ and $ D_i := z_i \frac{\partial}{\partial z_i} $ define
operators $ Y_i \in End( \Cz)  \iN $:
\bqa
Y_i & := & g_{i,i+1}^{-1}K_{i,i+1}\dots
g_{i,N}^{-1}K_{i,N}p^{D_i}K_{1,i}g_{1,i}\dots K_{i-1,i}g_{i-1,i}
\eqa
Taken together with $ { \gi }\; (i=1,\dots, N-1 )$  these operators satisfy the
relations of the Affine Hecke Algebra  $ \widehat{\hecke} $ [BGHP] :
\bqa
\gi^2 & = & (q-q^{-1})\gi + 1  \\
\gi g_{k,k+1} & = & g_{k,k+1} \gi \; , \qquad |i-k| \geq 2  \\
\gi g_{i+1,i+2}\gi & = & g_{i+1,i+2}\gi g_{i+1,i+2} \\
Y_k \gi & = & \gi Y_k \; , \qquad k \neq i,i+1 \\
 \gi Y_i & = & Y_{i+1} \gi^{-1}  \\
     Y_i Y_j & = & Y_j Y_i \: .
\eqa

Any symmetric polynomial in $ Y_i \; \iN $ belongs to the center of $
\widehat{\hecke} $. All symmetric polynomials in $ Y_i \; \iN $
are generated by the elementary symmetric polynomials which are obtained  by
expanding in the  parameter $u$ the generating function $ \Delta(u) $ :
\bqa
\Delta(u) & = & \prod_{\iN }(1 + uY_i)
\eqa

\subsubsection{The representation of the Hecke Algebra $ \hecke $ in $
(\cplx^2)^{\otimes N} $ }

Let $ V := \cplx^2 = {\rm span} \{ v^+,v^- \} $. Define $ t \in End(V\otimes V)
$ by
\bqa
t v^+\otimes v^- & = & (q-q^{-1})v^+\otimes v^- + v^-\otimes v^+ \\
t v^-\otimes v^+ & = & v^+\otimes v^-  \\
t  v^\pm\otimes v^\pm & = & qv^\pm\otimes v^\pm
\eqa

The operators $ \Pi^\pm(q) $:
\bqa
\Pi^+(q) & := & \frac{ q^{-1} +  t }{q + q^{-1} } \\
\Pi^-(q) & := & \frac{ q  -  t }{q + q^{-1} }
\eqa
are orthogonal projectors on the subspaces:
\bqa
S_q( V\otimes V ) & := & \cplx \{ v^+\otimes v^+ , v^-\otimes v^- , qv^+\otimes
v^- + v^-\otimes v^+ \} \\
A_q( V\otimes V ) & := & \cplx \{ v^+\otimes v^- - q v^-\otimes v^+ \}
\eqa
respectively.

Let $ H := V^{\otimes N}$. For an $ O \in  End(V\otimes V) $ denote by $
O_{i,j} \in End(H) $ the standard injection  $ End(V\otimes V) \rightarrow
End(H) $ which acts trivially on all the factors except the $i$-th and $j$-th
ones.

The matrices $ t_{i,i+1} \quad (i=1,\dots, N-1) $ satisfy the defining
relations (1.1.2-.4) of the finite-dimensional Hecke Algebra $ \hecke $.

\subsection{The Algebra $ \glt $ at level 0 and some of its  representations}
In this subsection we summarize several facts about the algebra $\glt$ and its
representations. Our conventions and notations mainly follow [JKKMP].

\subsubsection{The Algebra $ \glt $ }

In the $L$-operator formalism $ U \equiv \glt  $ at zero level is defined to be
the associative algebra with unit generated by
elements $ l_{ij}^\pm[\pm n]\;(i,j = 1,2 \:; n = 0,1,\dots) $.The $L$-operators
$ L^\pm(u) \in End(V\otimes U) $ are the generating series in the spectral
parameter $u$:
\bqa
L^\pm(u) & := & \sum_{\pm n \geq 0 } u^n \left( \begin{array}{c c}  l_{11}[n] &
l_{12}[n] \\
                                                                l_{21}[n] &
l_{22}[n]
                                                                  \end{array}
\right)
\eqa

The defining relations of $U$ are written in the the form:
\bqa
     \bar{R}_{ab}(u/v)L^\pm_a(u)L^\pm_b(v) & = &
L^\pm_b(v)L^\pm_a(u)\bar{R}_{ab}(u/v) \\
     \bar{R}_{ab}(u/v)L^+_a(u)L^-_b(v) & = & L^-_b(v)L^+_a(u)\bar{R}_{ab}(u/v)
\\
      l_{ii}^+[0]l_{ii}^-[0]\; = \; 1 \quad (i=1,2)\:, &  & l_{21}^+[0] \: = \:
l_{12}^-[0] \: = \: 0
\eqa
where the $R$-matrix $ \bar{R}_{ab}(z) \in End(V\otimes V := V_a \otimes V_b )
$ is defined as follows:
\bqa
         \bar{R}(z) & = &  \frac{ zt - t^{-1} }{qz - q^{-1}}P
\eqa
by $P$ we denote the permutation operator in $ V\otimes V $.

\subsubsection{Some representations of $U$}

A finite-dimensional highest weight module $W$ of $U$ contains non-zero vector
$\Omega$ which satisfies the condition:
\bqa
L^\pm(u)\Omega & = & \left( \begin{array}{c c}
                             A^\pm(u) &  *  \\
                             0 &   D^\pm(u)
                             \end{array} \right) \Omega
\eqa
where $ A^\pm(u)$ and $ D^\pm(u)$ are $\cplx$-valued series in $u$.

If $W$ is  irreducible, it is  specified up to equivalence by its Drinfeld
polynomial $Q(u)$ whch is determined by the conditions: $ Q(0) \: = \: 1 $ and
\bqa
        q^{degQ}\frac{Q(q^{-2}u)}{Q(u)} & = & \frac{A^+(u^{-1})}{D^+(u^{-1})}
\qquad ( u \rightarrow 0 ) \\
                                        & = & \frac{A^-(u^{-1})}{D^-(u^{-1})}
\qquad ( u \rightarrow \infty )
\eqa

The example of such $W$ is the 2-dimensional evaluation module $ W(a) $ where $
a \in \cplx \backslash \{0\}  $ is the parameter. As a vector space $W(a)$ is
isomorphic to $V$. The generators  $ l_{ij}^\pm[\pm n]\;(i,j = 1,2 ; n =
0,1,\dots) $ are defined by expanding the $L$-operator:
\bqa
L(ua) & := & \frac{uat - t^{-1}}{ua - 1}P \; \in End(V_a \otimes V)
\eqa
into power series in $ u^\pm $ around  zero and infinity respectively.

The Drinfeld polynomial of $W(a)$ is: $ Q(u;a) \; = \; 1 - a^{-1}u .$

A pair of tensor products $W(a)\otimes W(b)$, $W(b)\otimes W(a)$ is intertwined
by the matrix
\bqa
\bar{Y}(z) & = & zt - t^{-1} \quad ( z \in \cplx ) \; \in End(V\otimes V )
\eqa
i.e.:
\bqa
\bar{Y}_{12}(a/b)L_{a1}(ua)L_{a2}(ub) & = &
L_{a1}(ub)L_{a2}(ua)\bar{Y}_{12}(a/b)
\eqa
this relation holds in the tensor product of an auxiliary copy of $V$ indicated
by the subscript $a$ and $ V \otimes V $ indicated by subscripts 1 and 2.

The intertwiner $\bar{Y}(a/b)$ is invertible unless either $ a = q^2b $, in
which case
\bqa
\bar{Y}(q^2) & = & (q^2 - 1)(q+q^{-1}) \Pi^+(q)
\eqa
or $ a = q^{-2}b $, in which case
\bqa
\bar{Y}(q^{-2}) & = & (1-q^{-2})(q+q^{-1}) \Pi^-(q)
\eqa

Together with (26) this leads to the invariance relations:
\bqa
L_{a1}(q^2u)L_{a2}(u):A_q(V\otimes V) & \subset & A_q(V\otimes V) \\
L_{a1}(q^2u)L_{a2}(u):S_q(V\otimes V) & \subset & A_q(V\otimes V)\oplus
S_q(V\otimes V)\\
L_{a1}(u)L_{a2}(q^2u):S_q(V\otimes V) & \subset & S_q(V\otimes V) \\
L_{a1}(u)L_{a2}(q^2u):A_q(V\otimes V) & \subset & A_q(V\otimes V)\oplus
S_q(V\otimes V)
\eqa

\subsection{The hierarchy of $U$-invariant Dynamical Models}

The central elements of $\widehat{\hecke}$ generated by $\Delta(u)$ were
proposed in [BGHP] to define the hierarchy of integrable Dynamical Models which
are  trigonometric - that is $U$-invariant - generalizations of the
Yangian-invariant Dynamical Models found by the same authors.

Define $ T_a(u) \in End(\Cz\otimes H) $ by taking the tensor product of the
$L$-operators (24):
\bqa
T_a(u)&=& L_{a1}(uY_1)L_{a2}(uY_2)\dots L_{aN}(uY_N)
\eqa
After expansion in $u^\pm$, $ T_a(u) $ gives rise to a representation of $U$ in
$ {\cal P} \: := \: \Cz \otimes H $.
The action of $\Delta(u)$ naturally extends to $ {\cal P}$. Retain the same
notation $\Delta(u)$ for this extension.
The $U$-invariance and integrability relations for $\Delta(u)$ are immediate:
\bqa
 [ \Delta(u) , T_a(v) ] & = & 0 \\
 \mbox{} [ \Delta(u) , \Delta(v) ] & = & 0
\eqa

Both $\Delta(u)$ and $ T_a(v)$ act in the (``bosonic'') subspace $\B$:
\bqa
\B & := & \{ b \in {\cal P} |  \gi b = t_{i,i+1}b \; (i=1,\dots,N-1) \}
\eqa
This allows to restrict  $\Delta(u)$ and $ T_a(v)$ on $\B$ where both of these
operators can be rewritten in such a way that they do not depend explicitely on
the operators $K_{i,j}$.

\subsection{Eigenvalue spectrum of the operators $ Y_i \; \iN $ }

{\bf 1.} We shall work in the  monomial basis of $ \Cz $.

 Introduce a convenient parametrization of the elements of this basis. With a
monomial $ z^{\nu} := z_1^{\nu_1}z_2^{\nu_2}\dots z_N^{\nu_N} \quad (\nu_i \in
{\Bbb Z}^+ \iN ) $ associate a partition $ \l := (\l_1 \geq \l_2 \geq \dots
\geq \l_N \geq 0 ) $ such that $  \{\nu_1,\nu_2,\dots,\nu_N \} = \{
\l_{\s_1},\l_{\s_2},\dots,\l_{\s_N} \}:= \s.\{ \l_1,\l_2,\dots,\l_N \} $ for
some $  \s \in S_N = $  symmetric group with $ N-1 $   generators. A partition
$ \l $ is uniquely specified by $ \nu $, while in general $ \s $ is not.

For $ \s \in S_N : \{ \s_1,\s_2,\dots , \s_N \} = \s.\{1,2,\dots,N \} $ define
$  p^{\s}_i \; : \;  i = \s_{p^{\s}_i} \quad \iN .$ Let $ \Lambda_N $ be the
set of all {\it N-}member partitions. For any $\l \in \Lambda_N $ write: $\l =
(\l_1=\l_2=\dots =\l_{m_1} > \l_{m_1+1}=\l_{m_1+2}=\dots =\l_{m_2} >  \ldots
\\  \ldots  > \l_{m_M+1}=\l_{m_M+2}=\dots=\l_N )$.

\begin{df}
For any $ \l \in \Lambda_N $:
\begin{align*}
\S := \{ \s \in S_N \; | \; p^{\s}_1 < p^{\s}_2 < \dots p^{\s}_{m_1} , \;
 p^{\s}_{m_1+1} < p^{\s}_{m_1+2} < \dots p^{\s}_{m_2},  \ldots  \\ \ldots ,
p^{\s}_{m_M+1} < p^{\s}_{m_M+2} < \dots p^{\s}_N  \}.
\end{align*}
\end{df}

For a given $\l$ the elements of the set $ \S $ are in one-to-one
correspondence with distinct rearrangements of the sequence $ \{
\l_1,\l_2,\dots,\l_N \}.$ If $ \s \in \S $ then $  (i\,i+1)\s \in \S $ iff $
\l_{\s_i} \neq  \l_{\s_{i+1}} \quad (i=1,\dots,N-1) .$ The particular choice of
 $ \S $  as a subset of $ S_N $ parametrizing distinct rearrangements of a
partition will be explained by the Proposition 1.

Denote $ z^{\l_{\s}} := z_1^{\l_{\s_1}}z_2^{\l_{\s_2}}\dots z_N^{\l_{\s_N}}
\quad (\l \in \Lambda_N\:, \; \s \in \S ).$ With this notation:
\begin{eqnarray}
\Cz  =  \bigoplus_{ \l \in \Lambda_N } \oplus_{ \s \in \S } \cplx z^{\l_{\s}}.
\end{eqnarray}
{\bf 2.} Introduce an ordering on the set of monomials $ \{ \mon \} \quad ( \l
\in \Lambda_N,\; \s \in \S ) .$

We say that $ \l > \tilde{\l} \quad ( \l,\tilde{\l} \in \Lambda_N ) $ iff the
first ( counting from left ) non-vanishing element of the sequence $ \{ \l_1 -
\tilde{\l}_1, \l_2 - \tilde{\l}_2, \dots,  \l_N - \tilde{\l}_N \} $ is
positive. Fix $ \l \in \Lambda_N .$ We say that $ \s > \tilde{\s}\quad ( \s,
\tilde{\s} \in \S )$ iff the last non-vanishing element of the sequence $ \{
\l_{\s_1} - \l_{\tilde{\s}_1},  \l_{\s_2} - \l_{\tilde{\s}_2}, \dots  \l_{\s_N}
- \l_{\tilde{\s}_N} \}$ is negative.

For $ \l , \tilde{\l} \in \Lambda_N \:;\; \s \in \S , \: \tilde{\s} \in
S^{\tilde{\l}}_N $ define $ \l_{\s} > \tilde{\l}_{\tilde{\s}} $ iff either $ \l
> \tilde{\l} \:,$ or $ \l = \tilde{\l}\:, \s > \tilde{\s} $. The ordering on
monomials $ \mon $ is induced by the ordering on the exponents $ \l_{\s} \quad
( \l \in \Lambda_N,\; \s \in \S ) .$

\mbox{}

\noindent {\bf 3.} The action of $ \hecke $ in the monomial basis is found by a
straightforward computation to be as follows:
\begin{equation} \gij \mon = ( i < j ) =
\begin{cases}
(q-q^{-1}) \mon + qz^{\l_{(ij)\s}} + \text{``s.p.''} & \text{ if $\l_{\s_i} >
\l_{\s_j}$ }, \\
 q\mon & \text{ if  $\l_{\s_i} = \l_{\s_j}$ }, \\
 q^{-1}z^{\l_{(ij)\s}} + \text{``s.p.''} & \text{ if $\l_{\s_i} < \l_{\s_j}$ }.
\end{cases}
\end{equation}
And
\begin{equation} \gij^{-1} \mon = (i < j ) =
\begin{cases}
qz^{\l_{(ij)\s}} + \text{``s.p.''} & \text{ if $\l_{\s_i} > \l_{\s_j}$ },\\
 q^{-1}\mon & \text{ if  $\l_{\s_i} = \l_{\s_j}$ }, \\
(q-q^{-1}) \mon + q^{-1}z^{\l_{(ij)\s}} + \text{``s.p.''} & \text{ if
$\l_{\s_i} < \l_{\s_j}$ }.
\end{cases}
\end{equation}
Where ``s.p.'' means a linear combination of monomials with smaller partitions.

It follows that:
\begin{equation} K_{i,j}\gij \mon = ( i < j ) =
\begin{cases}
q\mon  + \text{``s.m.''} & \text{ if $\l_{\s_i} \geq \l_{\s_j}$ }, \\
q^{-1}\mon  + \text{``s.m.''} & \text{ if $\l_{\s_i} < \l_{\s_j}$ }.
\end{cases}
\end{equation}
And
\begin{equation} \gij^{-1}K_{i,j} \mon = ( i < j ) =
\begin{cases}
q^{-1}\mon  + \text{``s.m.''} & \text{ if $\l_{\s_i} \geq \l_{\s_j}$ }, \\
q\mon  + \text{``s.m.''} & \text{ if $\l_{\s_i} < \l_{\s_j}$ }.
\end{cases}
\end{equation}
Where ``s.m.'' signifies a linear combination of smaller monomials.

\mbox{}

\noindent {\bf 4.} The formulas of the preceding paragraph lead to the
following proposition:
\begin{prop}
The operators $ Y_i \; \iN $ are triangular in the monomial basis of $ \Cz $.
The action of these operators on monomials is given by:
\begin{equation*}
Y_i\mon = p^{\l_{\s_i}}q^{l_{\s_i}}\mon + \text{{\em``s.m.''}} \quad ( \l \in
\Lambda_N,\; \s \in \S,\; \iN )
\end{equation*}
where $ l_i := 2i-N-1 \quad \iN $.
\end{prop}
This proposition shows that $ \zeta_i^{\l}(\s) :=  p^{\l_{\s_i}}q^{l_{\s_i}}
\quad (\l \in \Lambda_N,\; \s \in \S) $ constitute a complete set of
characteristic numbers of the operator $ Y_i \quad (i \in \{ 1,2,\dots,N \}).$
In order to prove that the operators $ Y_i \; \iN $ are simultaneously
diagonalizable and that $ \{\zeta_i^{\l}(\s)\}\quad (\l \in \Lambda_N,\; \s \in
\S) $ form the complete set of eigenvalues of $ Y_i \quad (i \in \{ 1,2,\dots,N
\}) $ we shall make use of Lemma 1  discussed in the next paragraph.

\mbox{}

\noindent {\bf 5.}  The aim of this paragraph is to recall the following (
presumably well-known ) result:
\begin{lemma}
Let $ \cV = \cplx\{f_a\}_{a = 1,2,\dots,d={\mathrm {dim}}\cV} $ be a
finite-dimensional vector space. Let $ \cZ_i  \in  End(\cV)\quad (i=1,2,\dots ,
\cN) $ and :
\begin{align}
[\cZ_i,\cZ_j] &  =   0 \quad  (i,j = 1,2,\dots, \cN), \tag{a} \\
\cZ_i \quad (i=1,2,\dots , \cN) \quad &    \text{are simultaneously triangular
in the basis}\: \{f_a\}_{a = 1,2,\dots,d}: \notag \\
\cZ_i f_a &  =   \xi_i^a f_a + \sum_{b < a} m_i^{b\,a} f_b \quad  (i=1,2,\dots
, \cN), \tag{b}   \\
 & \text{where $ m_i^{b\,a} $ are coefficients.}  \notag
\end{align}
{\em (c)} The joint set of characteristic numbers $ \{ \xi_i^a \} \quad
(i=1,2,\dots,\cN \:;\; a = 1,2,\dots,d ) $ is multiplicity-free:
\begin{align}
\forall\; a \neq b \; ( a,b = 1,2,\dots,d )\; & \; \exists \; I(a,b) \subset
\{1,2,\dots,\cN\}:  \notag \\
\forall \; i \in I(a,b) \; & \xi_i^a - \xi_i^b \neq 0 . \notag
\end{align}

\noindent Then $ \exists $ a basis $ \{ \phi_a \}_{a=1,2,\dots,d} $ :
\begin{align}
\cZ_i \phi_a & = \xi_i^a \phi_a \quad ( a=1,2,\dots,d ;\: i=1,2,\dots,\cN )
\notag \\
\phi_a & = f_a + \sum_{b < a } \phi_{b\,a} f_b \quad ( a=1,2,\dots,d )  \notag
\end{align}
Where the coefficients $ \phi_{b\,a} $ are recursively defined as follows:
\begin{align}
\phi_{b\,a} & = \frac{1}{ \xi^a(w) - \xi^b(w) } ( m_{b\,a}(w) + \sum_{b < c <
a}m_{b\,c}(w)\phi_{c\,a}), \notag   \\
\text{here} \quad \xi^a(w)   & := \sum_{i=1}^{\cN}w^{i-1}\xi_i^a \,,  \qquad
 m_{a\,b}(w) \; := \;  \sum_{i=1}^{\cN}w^{i-1} m_i^{a\,b}\,.  \notag
\end{align}
$ w \in \cplx $ and $ \phi_{b\,a} $  does not depend on w.
\end{lemma}

\mbox{}

\noindent {\bf 6.} The joint characteristic number spectrum $ \{ \zeti =
p^{\l_{\s_i}}q^{l_{\s_i}} \}\quad  (\iN ; \; $ \mbox{} $ \l \in \Lambda_N ; \;
\s \in \S )  $ of the operators $ Y_i \quad \iN $ is explicitely
multiplicity-free.The operators $ \gij \quad ( i,j = 1,2,\dots,N-1 ) $ preserve
the finite-dimensional subspaces of $ \Cz $ formed by homogeneous polynomials
of any total degree. Therefore  $ \gij \quad ( i,j = 1,2,\dots,N-1 ) $ and
consequently  $ Y_i \quad \iN $ are direct sums of finite dimensional
operators. So we can apply the result of Lemma 1 and arrive at the following
proposition:
\begin{prop}
There exist polynomials $ \Phi^{\l}_{\s}\quad  (\l \in \Lambda_N ; \; \s \in \S
) $ s.t.:
\begin{align}
\Cz & = \bigoplus_{\l \in \Lambda_N}E^{\l} \; , \;
E^{\l}\::=\;\oplus_{\s \in \S} \cplx\Phi^{\l}_{\s} \, , \tag{i} \\
Y_i \Phi^{\l}_{\s} & = \zeti \Phi^{\l}_{\s} \quad \iN , \tag{ii} \\
\Delta(u) \Phi^{\l}_{\s} & =  \prod_{i=1}^N (1 + up^{\l_i}q^{l_i})
\Phi^{\l}_{\s} \,, \tag{ii'} \\
\Phi^{\l}_{\s} & = \mon + \text{{\em``s.m.''}}.  \tag{iii}
\end{align}
\end{prop}

\subsection{Action of the Hecke Algebra in the eigenspaces $E^{\l}\quad ( \l
\in \Lambda_N )$ of the operator $\D$.}

\noindent {\bf 1.} Fix any $ \l \in \Lambda_N $ . Since $[\widehat{\hecke}\:,\:
\D]\:=\: 0$ with the action of $\widehat{\hecke}$ defined in {\bf 1.1.1}, we
have:\begin{equation*}
 \widehat{\hecke}: \: E^{\l} \rightarrow E^{\l}
\end{equation*}
The affine generators of $ \widehat{\hecke} $ act in $ E^{\l} $ as given by
(ii) in Proposition 2. In this section we find the action of the
finite-dimensional Hecke Algebra generated by $ \gi \quad (i=1,...,N-1) $ on
the polynomials $\Ph \quad ( \s \in \S ) $ forming a basis in $ E^{\l} \quad (
\l \in \Lambda_N ). $

\noindent {\bf 2.} Introduce operators $ T_{i,i+1}\; \in \; \widehat{\hecke}
\quad ( \iNm ) $ :
\begin{equation}
T_{i,i+1} := \gi (Y_{i+1} - Y_i ) - (q-q^{-1}) Y_{i+1} \quad (\iNm).
\end{equation}
These operators satisfy the following relations:
\begin{align}
T_{i,i+1}Y_i \; = \; Y_{i+1} T_{i,i+1}\:, & \quad   T_{i,i+1}Y_{i+1} \; = \;
Y_i T_{i,i+1} \quad (\iNm)  \\  \mbox{}  [ T_{i,i+1}, Y_j ]& = 0  \qquad  ( j
\neq i,i+1)
\end{align}
Since $ T_{i,i+1}\; \in \; \widehat{\hecke} $, we have: $ T_{i,i+1}: E^{\l}
\rightarrow E^{\l} \quad ( \iNm ). $

\mbox{}

\noindent {\bf 3.} Consider the vector $ T_{i,i+1}\Ph \in E^{\l} \quad ( \s \in
\S \; , \; i \in {1,2,\dots,N-1} ). $ Due to (1.1.43,44):
\begin{eqnarray*}
Y_k T_{i,i+1}\Ph  & = & \xi_k^{\l}((i,i+1)\s) T_{i,i+1}\Ph  \quad ( k =
1,\dots,N ).
\end{eqnarray*}
Since the spectrum of $ Y_k \quad ( k = 1,\dots,N ) $ on $ E^{\l}$ does not
contain $ \xi_k^{\l}(\s) $ s.t. $ \s \not\in \S $, we have:
\begin{align}
 T_{i,i+1}\Ph & =   0 \quad \text{when $ \s \in \S\,, (i,i+1)\s \not\in \S $}
\quad (\iNm).
\end{align}
Since the joint spectrum of $ Y_i \; \iN $ on $ E^{\l}$ is multiplicity-free,
we have:
\begin{align}
T_{i,i+1}\Ph  & = \tau_{i,i+1}^{\l}(\s) \Phh \quad  \text{when $ \s , (i,i+1)\s
\in \S $} \quad (\iNm).
\end{align}
Where $\tau_{i,i+1}^{\l}(\s)$ is a coefficient.

Let $ \s \in \S $ and $ i $ be s.t. $ (i,i+1)\s \not\in \S $. Then $ \l_{\s_i}
= \l_{\s_{i+1}} \;,\; \s_{i+1} = \s_i + 1 $, and (1.1.45) gives:
\begin{align}
(\gi - q )\Ph & = 0 .
\end{align}
Let $ \s \in \S $ and $ i $ be s.t. $ (i,i+1)\s \in \S $. Then we can recast
(1.1.46) as follows:
\begin{align}
\gi\Ph & = \frac{(q-q^{-1})\zetii}{\zetii -\zeti}\Ph + \frac{
\tau_{i,i+1}^{\l}(\s)}{\zetii-\zeti}\Phh\, .
\end{align}

\noindent In order to find  $\tau_{i,i+1}^{\l}(\s)$  we shall equate
coefficients standing in front of monomials  $ \mon \, ,\; \moni $ in the both
sides of eq. (1.1.48). Recall that  $ \s ,\, (i,i+1)\s \in \S $ entails in
particular $ \l_{\s_i} \neq \l_{\s_{i+1}} $.

\noindent Let $ \l_{\s_i} > \l_{\s_{i+1}}\,\Rightarrow\, \mon > \moni $.

\noindent According to (1.1.38) the monomial $ \mon $ which is the maximal
monomial in $\Ph$ appears in  $\gi\Ph$ from two sources: from $\gi\mon$ and
from $\gi\moni$. Denote by $x$ the coefficient at the monomial $\moni$ in
$\Ph$.

\noindent In the RHS of (1.1.48) $\mon$ appears as the maximal monomial in
$\Ph$ and does not appear in $\Phh$.

Using (1.1.38) to compute the contributions from $\gi\Phi$ we equate the
coefficients in front of $\mon$ in the both sides of (1.1.48):
\begin{align}
\d + q^{-1}x & = \frac{\d\zetii}{\zetii - \zeti} .
\end{align}

Computing the contribution from $\moni$ we find that in the LHS of (1.1.48)
this monomial appears only in $\gi\mon$, while in the RHS it appears with
coefficient $x$ in $\Ph$ and as the maximal monomial in $\Phh$. Equating the
coefficients we get:
\begin{align}
q & = \frac{\d\zetii}{\zetii - \zeti} x + \frac{\tau_{i,i+1}^{\l}(\s)}{\zetii -
\zeti} \,.
\end{align}
Combining (1.1.49) and (1.1.50) we obtain:
\begin{align}
\tau_{i,i+1}^{\l}(\s) & = q\frac{(q^{-1}\zetii - q\zeti)(q\zetii -
q^{-1}\zeti)}{\zetii - \zeti} \quad ( \l_{\s_i} > \l_{\s_{i+1}})
\end{align}

\mbox{}

\noindent   Let $ \l_{\s_i} < \l_{\s_{i+1}}\,\Rightarrow\, \mon < \moni $.
Equate the coefficients in front of the monomial $\moni$ in the both sides of
(1.1.48).

\noindent In the LHS the contribution comes only from $\gi\mon$. In the RHS
only $\Phh$ contributes $\moni$ as its maximal monomial. Application of
(1.1.38) leads to:
\begin{align}
q^{-1} & =  \frac{\tau_{i,i+1}^{\l}(\s)}{\zetii - \zeti} \quad  ( \l_{\s_i} <
\l_{\s_{i+1}}).
\end{align}

\mbox{}

\noindent {\bf 4.} To summarize, we have obtained the following proposition:
\begin{prop}
Let $ \s \in \S \quad (\l \in \Lambda_N)$, then $\hecke$ acts in
$E^{\l}\:=\:\cplx\{\Ph \}_{\s \in \S}$ as follows: \begin{equation} \gi\Ph =
\frac{(q-q^{-1})\zetii}{\zetii -\zeti}\Ph +
\begin{cases}
q\frac{(q^{-1}\zetii - q\zeti)(q\zetii - q^{-1}\zeti)}{(\zetii - \zeti)(\zetii
- \zeti)}\Phh &  \\ \text{{\em when} $\l_{\s_i} > \l_{\s_{i+1}}\,\Rightarrow\,
(i,i+1)\s \in \S $}, &   \\
0 \quad  \text{{\em when} $\l_{\s_i} = \l_{\s_{i+1}}\,\Leftrightarrow\,
(i,i+1)\s \not\in \S $}, &  \\
q^{-1}\Phh &  \\  \text{{\em when} $\l_{\s_i} < \l_{\s_{i+1}}\,\Rightarrow\,
(i,i+1)\s \in \S $}. &
\end{cases}
\end{equation}
\end{prop}

\section{The limit $p \rightarrow 1$ of the hierarchy of Dynamical Models.}

\subsection{Few facts about Macdonald operators.}

{\bf 1.}The Macdonald operators $ D_N^n(p,t) \quad (n=0,\dots,N)$ [M] act in
the subspace of $\Cz$ formed by symmetric polynomials. In notation of [JKKMP]
these operators are defined as follows:
\begin{equation}
D_N^n(p,t) := t^{n(n-1)/2}\sum_{I_n}\prod\begin{Sb}
                                     i\in I_n \\
                                     j\not\in I_n
                                            \end{Sb}
\frac{tz_i - z_j}{z_i - z_j}\prod_{k\in I_n} p^{D_k} \quad (n=0,\dots,N),
\end{equation}
where the summation is over all subsets $ I_n $ of $\setN$ which contain $n$
elements. Using the formula:
\begin{equation*}
det\left\| \frac{(t-1)w_i}{tw_i-w_j}\right\|_{1\leq i,j\leq m} = \quad
t^{m(m-1)/2}\prod_{1\leq i\neq j\leq m}\frac{w_i-w_j}{tw_i-w_j},
\end{equation*}
where $w_i \quad (i=1,\dots,m)$ are numbers; we can rewrite the definition of
the operators $ D_N^n(p,t) \quad (n=0,\dots,N)$ in another form:
\begin{equation}
D_N^n(p,t) = \sum_{I_n}detA_{I_n}(t)\prod_{k\in I_n} p^{D_k} \quad
(n=0,\dots,N),
\end{equation}
where $A_{I_n}(t)$ is a submatrix of the matrix:
\begin{equation}
A(t) = \| A_{ij}(t)\|_{1\leq i,j\leq N}\;,\;A_{ij}(t):=\frac{(t-1)z_i}{tz_i -
z_j} \prod\begin{Sb}
                                     1\leq k\leq N \\
                                     k\neq i
                                            \end{Sb}\frac{tz_i - z_k}{z_i -
z_k} ;
\end{equation}
which is defined as follows:$ A_{I_n}(t):=\| A_{ij}(t)\|_{i,j\in I_n}.$

The Macdonald polynomials $P_{\l}(p,t) \quad (\l \in \Lambda_N)$ are
eigenfunctions of the operators $ D_N^n(p,t) \quad (n=0,\dots,N)$:
\begin{equation}
D(v;p,t)P_{\l}(p,t) = \prod_{i=1}^N (1 + t^{N-i}p^{\l_i}v)\, P_{\l}(p,t) \quad
(\l \in \Lambda_N),
\end{equation}
here $D(v;p,t)$ is the generating function of Macdonald operators:
\begin{equation*}
D(v;p,t):= \sum_{n=0}^N v^n  D_N^n(p,t) \;.
\end{equation*}

\noindent {\bf 2.} In the limit $p \rightarrow 1$ one finds [M]:
\begin{equation}
P_{\l}(p,t) = e_{\l '} + O(p-1)  \quad (\l \in \Lambda_N),
\end{equation}
where $\l '$ is the conjugate partition of $\l$ and for a partition $\pi \: :
(\pi_1 \geq  \pi_2 \geq \dots ) \; , \; (\pi_1 \leq N) \quad e_{\pi}:=
e_{\pi_1}e_{\pi_2}\ldots \quad ;$ where $ e_r $ is the elementary symmetric
polynomial:
\begin{equation*}
e_r := \sum_{1\leq i_1 \leq\dots\leq i_r\leq N} z_{i_1}z_{i_2}\ldots z_{i_r}.
\end{equation*}

Consider the limit $p \rightarrow 1$ of the generating function $D(v;p,t)$:
\begin{equation}
D(v;p,t) \overset{p \rightarrow 1}{=} D_0(v;t) + (p-1) D_1(v;t) + O((p-1)^2).
\end{equation}
{}From (2.2.1,.4,.5) it follows that $D_0(v;t)$ is a multiplication by a
constant:
\begin{equation}
D_0(v;t) = \prod_{i=1}^N (1 + t^{N-i}v) \;.
\end{equation}
The first-order term $D_1(v;t)$ is a differential operator:
\begin{equation}
D_1(v;t) = \sum_{i=1}^N \left(\sum_{n=1}^N v^n \sum_{I_n: i\in I_n}
detA_{I_n}(t)\right)D_i
\end{equation}
Expanding the eq. (2.2.4) up to the first order in $ p-1 $ and using (2.2.5,.7
) we get:
\begin{equation}
D_1(v;t)e_{\l '} = \left(v\sum_{j=1}^N\prod\begin{Sb} 1\leq k\leq N \\
                        k\neq j \end{Sb} (1 + v t^{N-k})
t^{N-j}\l_j\right)\,e_{\l '} \quad ( \l \in \Lambda_N ).
\end{equation}

\subsection{Taking the limit  $p \rightarrow 1$ in the hierarchy of Dynamical
Models}

{\bf 1.} The following fact was established in the paper [JKKMP]. Let $S$ be
any symmetric polynomial $ ( S \in \Cz )$. The action of the operator $\D$ (cf.
{\bf 1.1}) on such $S$ coincides with the action of the generating function of
Macdonald operators:
\begin{equation}
\D S = D(q^{N-1}u;p,q^{-2})S \;.
\end{equation}

There is another connection between $\D$ and Macdonald operators.
For an operator $O$ which is a function of operators $ D_i , z_i \; \iN ,
K_{i,i+1} $ $ (i = 1,\dots,N-1) $ introduce a normal ordering $:\; :$. The
normal ordering is described as follows: in $O$ bring all the operators $ D_i $
to the right {\em without} taking commutators between  $ D_i $ and $ z_j $, but
taking into account the commutation relations between $ D_i $ and $  K_{j,j+1}
$. For instance:
\begin{equation*}
 :\; p^{D_1}\, \frac{q^{-1}z_1 - qz_2}{z_1-z_2}(K_{12}  - 1)\; : \;=
\frac{q^{-1}z_1 - qz_2}{z_1-z_2}(K_{12} \,p^{D_2}- p^{D_1}) .
\end{equation*}
\noindent We formulate the following Lemma:
\begin{lemma}
\begin{equation}
 :\; \D \; : \; =  D(q^{N-1}u;p,q^{-2}) \; .
\end{equation}
\end{lemma}
\begin{pf}
To facilitate the proof we introduce an extension of the algebra generated by $
z_1,z_2,\dots,z_N $ and $ K_{1,2},K_{2,3},\dots,K_{N-1,N}$ by symbols $
\xi_1,\xi_2,\dots,\xi_N $. These symbols are defined by the commutation
relations
\begin{align*}
[\xi_i,\xi_j] & = 0  \; , \; [\xi_i,z_j] = 0 \qquad ( i,j \in \setN), \\
K_{i,j}\xi_j & = \xi_i K_{i,j}\;,\; [K_{i,j},\xi_k] = 0 \quad (k\neq i,j)
\qquad ( i\neq j \in \setN).
\end{align*}
One can take $\xi_i := f(z_i) $, where $f$ is any function of one variable, as
a realization for $ \xi_i$. \\
Together with $ \gi $ the operators
\begin{equation*}
Y_i(\xi) := g_{i,i+1}^{-1}K_{i,i+1}\dots g_{i,N}^{-1}K_{i,N}\xi_i
K_{1,i}g_{1,i}\dots K_{i-1,i}g_{i-1,i}
\end{equation*}
still satisfy the defining relations of $ \widehat{\hecke}$. \\
This implies in particular that for the operator
\begin{equation*}
\Delta(u;\xi) := \prod_{i=1}^N(1 + u Y_i(\xi))
\end{equation*}
we have
\begin{equation}
[\Delta(u;\xi),\gi] = 0 \qquad \iNN. \tag{i}
\end{equation}
Using the commutation relations with $ z_1,z_2,\dots,z_N $ and $ K_{i,j} $ we
can bring the symbols $ \xi_i $ to the right of all expressions in
$\Delta(u;\xi)$. Denote $\Delta(u;\xi)$ with all $ \xi_i $ brought to the right
by  $\Delta(u;\xi)'.$ \\
We have
\begin{equation*}
: \Delta(u) : = \Delta(u;\xi)' |_{\xi_i \rightarrow p^{D_i}}.
\end{equation*}
Therefore in order to prove the statement of the lemma we compute the
coefficients standing in front of monomials $ \xi_{i_1}\xi_{i_2}\dots \xi_{i_n}
 \quad (1\leq i_1 \leq i_2 \leq \dots \leq i_n \leq N)$ in the symmetric
functions
\begin{equation*}
\Delta(\xi)^{(n) } := \sum_{N \geq k_1 > k_2 > \dots > k_n \geq 1 }
Y_{k_1}(\xi)Y_{k_2}(\xi)\dots  Y_{k_n}(\xi) \quad ( 1 \leq n \leq N ).
\end{equation*}
With notation of {\em {\bf 1.1.1}} we have
\begin{align*}
r_{i,j} & := K_{i,j}g_{i,j} = a_{j,i} + b_{j,i}K_{i,j} , \\
r_{i,j}^{-1} & = a_{i,j} - b_{j,i}K_{i,j}, \\
Y_i(\xi) & = r_{i,i+1}^{-1}\dots r_{i,N}^{-1}\xi_i r_{1,i}\dots r_{i-1,i}.
\end{align*}

\mbox{}

\noindent Let us compute the terms in $\Delta(\xi)^{(n)'} \quad ( 1\leq n \leq
N)$ which contain symbols $ \xi_1 ,\xi_2, \dots ,\xi_n $ only. By inspection we
find that such terms can appear only in $ Y_n(\xi)Y_{n-1}(\xi)\dots  Y_1(\xi)
$. Furthermore the relevant contributions from the individual factors in the
last expression are
\begin{align*}
Y_1(\xi) & \rightarrow  r_{1,2}^{-1}\dots r_{1,n}^{-1}a_{1,n+1}\dots a_{1,N}
\xi_1 , \\
Y_2(\xi) & \rightarrow  r_{2,3}^{-1}\dots r_{2,n}^{-1}a_{2,n+1}\dots a_{2,N}
\xi_2  r_{1,2}, \\
         &  \vdots  \\
Y_k(\xi) & \rightarrow  r_{k,k+1}^{-1}\dots r_{k,n}^{-1}a_{k,n+1}\dots a_{k,N}
\xi_k  r_{1,k}\dots r_{k-1,k}, \\
          & \vdots \\
Y_{n-1}(\xi) & \rightarrow  r_{n-1,n}^{-1}a_{n-1,n+1}\dots a_{n-1,N} \xi_{n-1}
r_{1,n-1}\dots r_{n-2,n-1}, \\
Y_n(\xi) & \rightarrow  a_{n,n+1}\dots a_{n,N} \xi_n  r_{1,n}\dots r_{n-1,n}.
\end{align*}
Multiplying these contributions we find that there is only one term in
$\Delta(\xi)^{(n)'}$ which contains  $ \xi_1 ,\xi_2, \dots ,\xi_n $ only; and
this term is
\begin{multline*}
(a_{1,n+1}\dots a_{1,N})(a_{2,n+1}\dots a_{2,N})\ldots (a_{n,n+1}\dots
a_{n,N})\xi_1\xi_2\dots\xi_n .
\end{multline*}

Next we use the Hecke-invariance relation {\em (i) } and find
\begin{equation*}
\Delta(\xi)^{(n)} = \sum_{I_n} \left(\prod\begin{Sb} i \in I_n \\ j \not\in I_n
\end{Sb} a_{i,j}\right) \prod_{i \in I_n } \xi_i.
\end{equation*}
Where the summation is over all $n$-element subsets of $\setN$. Comparing this
expression with {\em (2.2.1)} we obtain the statement of the lemma.
\end{pf}

\mbox{}

\noindent {\bf 2.} Let us take the limit $p \rightarrow 1$ in the operators $
\D , T_a(u), Y_i \quad \iN.$ Expanding around $ p = 1 $ and keeping the first
two terms of the expansion we write:
\begin{align}
Y_i  &\overset{p \rightarrow 1}{=} y_i + (p-1)x_i + O((p-1)^2)\quad (\iN),  \\
\D  &\overset{p \rightarrow 1}{=} \Delta_0(u) + (p-1)\Delta_1(u)+ O((p-1)^2)\:,
\\
T_a(u) &\overset{p \rightarrow 1}{=} T_a^0(u) + (p-1) T_a^1(u)+ O((p-1)^2)\: .
\end{align}
Here we introduced the operators:
\begin{align}
y_i & := g_{i,i+1}^{-1}K_{i,i+1}\dots g_{i,N}^{-1}K_{i,N} K_{1,i}g_{1,i}\dots
K_{i-1,i}g_{i-1,i} \quad (\iN) , \\
x_i & := g_{i,i+1}^{-1}K_{i,i+1}\dots g_{i,N}^{-1}K_{i,N}D_i
K_{1,i}g_{1,i}\dots K_{i-1,i}g_{i-1,i} \quad (\iN) , \\
& \Delta_0(u)  := \prod_{i=1}^N(1 + uy_i) \;, \\
& \Delta_1(u)  := u\sum_{j=1}^N \prod_{1\leq i<j}(1 + uy_i)\, x_j\prod_{j<k\leq
N}(1 + uy_k)\;,\\
T_a^0(u) & := L_{a1}(uy_1)L_{a2}(uy_2)\ldots L_{aN}(uy_N) \quad \in
End\Cz\otimes H\;.
\end{align}
The operators $y_i , g_{i,i+1}$ satisfy the relations (cf. {\bf 1.1}) of Affine
Hecke Algebra and $ T_a^0(u)$ defines a representation of $U$.

\noindent {\bf 3.} From  Lemma 2 it follows that the operators $ \Delta_0(u)\,
, \, \Delta_1(u) $ have  a rather special form. We have the proposition:
\begin{prop}
The following statements hold:

\noindent Let $D_0(v;t)\, , \; D_1(v;t)$ be those defined in {\em (2.2.6,.8)};
then:
\begin{align}
\Delta_0(u) & = D_0(q^{N-1}u;q^{-2}) \, = \, \prod_{i=1}^N (1 + u q^{2i-N-1})I
\, . \tag{i}
\end{align}
I.e. $\Delta_0(u)$ is a multiplication by a constant.
\begin{align}
\Delta_1(u) & =  D_1(q^{N-1}u;q^{-2}) + \Xi(u) \,. \tag{ii}
\end{align}
Where operator $ \Xi(u)$ is a function of operators $ z_i \,,\, K_{i,j} \quad
(i,j = 1,\dots, N) $ only ( and {\em not} of $ D_i $).
\end{prop}

\mbox{}

\noindent Let $ \D_1 := \sum_{i=1}^{N}u^i \Delta_1^{(i)} $ and $ \Xi(u) :=
\sum_{i=1}^{N}u^i \Xi^{(i)} $. We have computed explicit expressions for the
operators $ \Delta_1^{(N)} $ and $ \Delta_1^{(1)} $. In notation of section
{\bf 1.1} one has:
\begin{align}
\Delta_1^{(N)} & = D_1 + D_2 + \dots + D_N \, , \\
\Delta_1^{(1)} & = \sum_{i=1}^N \left( \prod\begin{Sb} 1\leq k\leq N \\ k\neq i
\end{Sb} a_{i,k} \right)D_i \, + \, \Xi^{(1)}\,
\end{align}
where:
\begin{multline}
\Xi^{(1)}  = \\ \sum_{M=2}^N \frac{(-1)^M}{\d}\sum_{N\geq i_M > \dots >i_1\geq
1} {\cal A}_{i_M,i_{M-1},\dots,i_1}{\cal B}_{i_M,i_{M-1},\dots,i_1}
K_{i_M,i_{M-1}}\dots K_{i_2,i_1} \; + \; \varphi^{(1)} \:, \notag \\
{\cal A}_{i_M,i_{M-1},\dots,i_1}  = \left(\prod_{i_1<f<i_2}
a_{i_1,f}\right)\left(\prod_{i_2<f<i_3} a_{i_2,f}\right)\ldots
\left(\prod_{i_M<f<N+i_1} a_{i_M,f\pmod{N}}\right) \:, \\
{\cal B}_{i_M,i_{M-1},\dots,i_1}  = b_{i_M,i_{M-1}}b_{i_M-1,i_{M-2}}\ldots
b_{i_2,i_{1}}b_{i_1,i_M} \:, \\
 \varphi^{(1)}  = -\sum_{1\leq k< i\leq N}\frac{a_{i,i+1}\dots a_{i,N}
a_{i,1}\dots a_{i,k-1}a_{i,k+1}\dots a_{i,i-1} b_{k,i} b_{i,k}}{\d} \: . \notag
\end{multline}

\mbox{}

\noindent {\bf 4.} Let us fix the notation:
\begin{equation}
{\cal D}(u) :=  D_1(q^{N-1}u;q^{-2}) = \sum_{i=1}^N \theta_i(u)\,D_i\, .
\end{equation}
Where according to (2.2.8) :
\begin{equation}
\theta_i(u) := \sum_{n=1}^N u^n q^{n(N-1)}\sum_{I_n : i\in I_n} det
A_{I_n}(q^{-2}) \quad (\iN).
\end{equation}
(Cf. sec. {\bf 2.1} for the definition of $A_{I_n}(t)$).
The functions $\theta_i(u)\quad (\iN)$ can be written in the following form:
\begin{equation}
\theta_i(u) = \frac{\partial}{\partial \gamma_i}det(I + u\Gamma q^{N-1}
A(q^{-2}))|_{\Gamma = I} \, ,
\end{equation}
where we have introduced an auxiliary matrix: $ \Gamma := {\mathrm
{diag}}\{\gamma_1,\ldots, \gamma_N \} .$
Using this representation we compute $\theta_i(u)\quad (\iN)$ at the point $
z_1 = \omega^1,\dots,z_N = \omega^N $, where $\omega := {\mathrm {exp}}(2\pi
i/N).$ The computation yields:
\begin{gather}
\theta_i(u)|_{z_1=\omega^1,\dots,z_N=\omega^N} = \frac{1}{N}\prod_{k=1}^N
(1+uq^{l_k})\,\sum_{n=1}^N \frac{uq^{l_n}}{1+uq^{l_n}} \; \equiv \theta(u)
\quad \iN. \\
l_i := 2i - N - 1  \notag
\end{gather}
Thus the point $ z_1 = \omega^1,\dots,z_N = \omega^N $ (or any point obtained
from it by a permutation of coordinates) is special in that at this point all
the $\{z_i\}$-dependent coefficients  $\theta_i(u)$ of the first-order
differential operator ${\cal D}(u)$ become equal one to another.

\section{Definition of the Hierarchy of Integrable, $U$-invariant Spin Models}

\subsection{Preliminaries}

{\bf 1.} Let us expand the relations (1.1.34,.35) around the point $p=1$ using
the definitions (2.2.12-.14) and the fact that $\Delta_0(u)$ is a constant. For
$\Delta_1(u),\,T_a^0(v) \in End(\Cz\otimes H)$ we obtain:
\begin{align}
 [\D_1,T_a^0(v)] & =  0 \; , \\
 \mbox{} [\Delta_1(u),\Delta_1(v)] & = 0.
\end{align}
Expanding the relations:
\begin{equation}
[\D,Y_i] = 0 \quad \iN,\quad [\D,g_{i,i+1}] = 0 \quad (i=1,\dots,N-1),
\end{equation}
we get:
\begin{equation}
[\Delta_1(u),y_i] = 0 \quad \iN,\quad [\Delta_1(u),g_{i,i+1}] = 0 \quad
(i=1,\dots,N-1).
\end{equation}
Due to (3.3.4) and the fact that $y_i \quad \iN ,\; g_{i,i+1} \quad
(i=1,\dots,N-1)$ satisfy the Affine Hecke Algebra relations, the operators
$\Delta_1(u), T_a^0(v)$ act in the ``bosonic'' subspace $\B$ (cf. 1.1.36 and
3.3.7).

\mbox{}

\noindent {\bf 2.} Introduce several definitions.
Let
\begin{align}
{\cal R} & := \cplx [\{\frac{1}{z_i - z_j}\}_{1\leq i\neq j\leq N}\,
,\{\frac{1}{z_i - q^{2}z_j}\}_{1\leq i\neq j\leq N}\,,\: z_1,\dots,z_N ]\otimes
H.
\end{align}
Let ${\cal P}$ and $\B$ be the subspaces of $ {\cal R}$:
\begin{align}
{\cal P} & := \Cz\otimes H  \, , \\
\B & := \{b\in {\cal P} | (\gi - t_{i,i+1})b = 0 \quad (i=1,\dots,N-1)\}.
\end{align}
For $v_1,\dots,v_N \, \in \cplx $ such that $ v_i \neq v_j\,,\,q^{2} v_j \quad
(i\neq j\,; i,j =1,\dots,N)$
define the evaluation map: $ Ev(v): {\cal R} \mapsto H $ by taking values of
rational functions at the point $
z_1 = v_1,\dots, z_N = v_N.\; ( \Leftrightarrow z = v ) $.

\noindent For any $ O \in End({\cal R}) $ define an operator $ \hat{O} \in
End({\cal R}) $ by the rule [BGHP]:

Using the commutation relations bring all the permutation operators
$K_{i,j}\quad (i,j \in \setN)$
to the right of an expression in $O$; replace the rightmost of $K_{i,j}$ using
the substitution:
\begin{equation}
 \gi \rightarrow t_{i,i+1}  \Rightarrow K_{i,i+1} \rightarrow
\frac{z_it_{i,i+1}^{-1} - z_{i+1}t_{i,i+1}}{q^{-1}z_i - qz_{i+1}} \quad
(i=1,\dots,N-1).
\end{equation}
Repeat the procedure until there are no operators  $K_{i,j}$ left. The result
is  $ \hat{O}$.

In what follows we  adopt the following notational convention: if ${\cal L}$ is
a linear space and $A,B$ are linear operators defined on  ${\cal L}$, we write:
\begin{gather*}
 A{\cal L} = B{\cal L}\quad  \text{meaning } \quad Al = Bl \quad \forall l \in
{\cal L}.
\end{gather*}
In particular for $O , \hat{O} \quad \in End({\cal R})$ defined above  we have:
\begin{equation}
O\B =  \hat{O}\B.
\end{equation}
Let $O,O' \in End({\cal P})$ be s.t.:
\begin{gather}
 [O,O']{\cal P} = 0 \quad \text{and}\quad  O ,O': \B\mapsto \B,
\end{gather}
then
\begin{equation}
[\hat{O},\hat{O'}]{\cal B} = 0.
\end{equation}

\noindent {\bf 3.}With notation of (2.2.20,.23,.26) let us consider the
following differential operator $\widetilde{{\cal D}(u)} \in End({\cal R})$:
\begin{equation}
\widetilde{{\cal D}(u)} := {\cal D}(u) - \theta(u)\Delta_1^{(N)}.
\end{equation}
Let $Ev(\omega)$ be the evaluation map $Ev(v)$ taken at the special point $ v_1
= \omega^1,\dots,v_N = \omega^N \quad ( \Leftrightarrow v = \omega ) $. Then in
virtue of (2.2.26) we obtain the following property of $ \widetilde{{\cal
D}(u)} $:
\begin{equation}
Ev(\omega)\widetilde{{\cal D}(u)}{\cal R} = 0.
\end{equation}
Let us introduce the modified generating function $ \widetilde{\Du} $ by
subtracting the product of the constant $\theta(u)$ and the operator
$\Delta_1^{(N)}$:
\begin{equation}
\widetilde{\Du} := \Du  - \theta(u)\Delta_1^{(N)} = \widetilde{{\cal D}(u)} +
\Xu .
\end{equation}
Since $\Delta_1^{(N)}$ is a member of the hierarchy of commuting operators
defined by $\Du$, the equations (3.3.1,.2),(3.3.4) still hold if we replace in
these equations $\Du$ by $\widetilde{\Du}$:
\begin{align}
[\widetilde{\Du},\widetilde{\Dv}]{\cal P} & = 0   ,  \\
[\widetilde{\Du}, T_a^0(v) ]{\cal P} & = 0 , \\
\widetilde{\Du} : {\cal B} &  \mapsto {\cal B} .
\end{align}

\subsection{Definition of the hierarchy of Spin Models}

{\bf 1.} Let $ H_{\B}(\omega)$ be the image of $\B$ under the action of the
evaluation map $Ev(\omega)$:
\begin{equation}
  Ev(\omega)\B =  H_{\B}(\omega) \subset H .
\end{equation}
Since $ \widehat{T_a^0(u)}\;,\,\widehat{\Xu} $ do not depend on the diffrential
operators $ D_i \; \iN $ and the operators of coordinate permutation, we can
define the operators $\Tuo , \Xuo$ as follows:
\begin{equation}
Ev(\omega)\widehat{T_a^0(u)} = \Tuo Ev(\omega) , \quad
Ev(\omega)\widehat{\Xi(u)} = \Xuo Ev(\omega).
\end{equation}
Applying $Ev(\omega)$ to the relations :
\begin{align}
\widehat{T_a^0(u)}:\B & \mapsto \B , \\
\widehat{\widetilde{\Du}}:\B & \mapsto \B,
\end{align}
and using (3.3.13) we get:
\begin{align}
\Tuo : H_{\B}(\omega) & \mapsto H_{\B}(\omega) , \\
\Xuo : H_{\B}(\omega) & \mapsto H_{\B}(\omega).
\end{align}

\noindent {\bf 2.} Apply the evaluation map $Ev(\omega)$ to the relations
(3.3.15,.16) taking (3.3.13) and (3.3.23) into account. As the result we find
that the operator $\Xuo$ is a generating function of the commuting,
$U$-invariant  integrals of motion which are operators in $ H_{\B}(\omega)$:
\begin{gather}
[\Xuo, \Xvo ] H_{\B}(\omega)  = 0 \, , \\
[\Xuo , \Tvo ]  H_{\B}(\omega)   = 0 , \\
\left(\bar{R}_{ab}(u/v)\Tuo T_b^0(v;\omega) -
T_b^0(v;\omega)\Tuo\bar{R}_{ab}(u/v)\right) H_{\B}(\omega) = 0 .
\end{gather}
In sec. 6 we shall show that $H_{\B}(\omega)=H$. This completes  the definition
of the hierarchy $ \Xi(\omega)^{(1)},\dots,\Xi(\omega)^{(N-1)} \; ( \Xuo =
\sum_{n=1}^{N-1} u^n \Xi^{(n)}(\omega) ) $  of Spin Models.

\section{Eigenvalue spectrum of the operators  $\Du\;,
\protect\newline y_i \; \iN $}

\subsection{Characteristic numbers and eigenvalues of  $\Du\;,\;y_i $}

{\bf 1.} To find the action of the operators  $\Du\;,\;y_i \; \iN$ in the
monomial basis of $\Cz$ we can take the limit $p\rightarrow 1$ in the formulas
of Proposition 1. This gives the following proposition:
\begin{prop}
The operators  $\Du\;,\;y_i \; \iN$ are triangular in the monomial basis of
$\Cz$. The action of these operators on monomials is given by:
\begin{align}
y_i\mon & = q^{l_{\s_i}}\mon + \text{{\em``s.m''}}\quad (\l \in \Lambda_n, \s
\in \S, \; \iN ), \tag{i} \\
\Du\mon & = \delta^{\l}(u)\mon + \text{{\em``s.m''}}\quad (\l \in \Lambda_N, \s
\in \S ) \tag{ii}.
\end{align}
where $l_i := 2i - N -1 \quad \iN $ and
\begin{equation*}
 \delta^{\l}(u) :=  u\sum_{j=1}^N\left(\prod\begin{Sb} 1\leq k\leq N \\ k\neq j
\end{Sb}(1 + uq^{l_k})\right)q^{l_j}\l_j .
\end{equation*}
\end{prop}

Since  $\Du\;$ and $\;y_i \; \iN$ commute among themselves (cf. 3.3.4) and the
joint spectrum of characteristic numbers of   $\Du\;$ and $\;y_i \; \iN$ given
by Proposition 5  is explicitely multiplicity-free, we  apply Lemma 1 and claim
that  $\Du\;,\;y_i \; \iN$ are simultaneously diagonalizable:
\begin{prop} There exist polynomials $\ph\quad (\l \in \Lambda_N, \s \in \S )$
s.t.:
\begin{align}
\Cz & = {\bigoplus}_{\l \in \Lambda_N}{\cal E}^{\l}\;,\;{\cal
E}^{\l}:={\oplus}_{\s\in\S}\cplx\ph\,,\tag{i}\\
y_i\ph & = q^{l_{\s_i}}\ph \quad \iN ,\tag{ii}\\
\Du\ph & = \delta^{\l}(u)\ph\,,\tag{ii'}\\
\ph & = \mon + \text{{\em``s.m''}}\,.
\end{align}
\end{prop}

\mbox{}

\noindent {\bf 2.} Let us show that $ \ph = {\lim}_{p\rightarrow 1} \Ph\quad
(\l \in \Lambda_N, \s \in \S )$ where $\Ph\quad (\l \in \Lambda_N, \s \in \S )$
are eigenfunctions of the operators $Y_i\quad\iN$ (Cf. Proposition 2).

Expanding $\Ph\quad (\l \in \Lambda_N, \s \in \S )$ around $p = 1$ let us
write:
\begin{equation}
\Ph \overset{p\rightarrow 1}{=} (p - 1)^{s} \psi^{\l}_{\s} + O((p-1)^{s+1}),
\end{equation}
where $ \psi^{\l}_{\s}$ is a non-zero polynomial. Since $ \Ph = \mon +
\text{{\em `` s.m.''}}
$, we have: $s \leq 0$.

Let us show that $s=0$. Suppose $s<0$. The satement (c) of Lemma 1 when applied
to the eigenvectors $\Ph$ enables us to detect which coefficients in the
decomposition  of $\Ph$ into monomials are potentially singular in the limit $p
\rightarrow 1$. The singularities may arise because of the presence of
denominators of the form
\begin{equation}
 \frac{1}{p^{\l_{\s_i}}q^{l_{\s_i}} - p^{\mu_{\s_i}}q^{l_{\s_i}}}
\end{equation}
where $\mu$ is a partition {\em smaller} than $\l$ and $\s \in \S , S^{\mu}_N
$.Therefore if $s$ in (4.4.2) is negative, the maximal monomial in $
\psi^{\l}_{\s}$ is {\em smaller } than any of the monomials $\mon \quad (\s \in
\S)$.

Expanding the equations (ii),(ii') of the Proposition 2 around the point $p=1$
and taking into account that $\Delta_0(u) = \D |_{p=1}$ is a constant, we
arrive at the following equations:
\begin{align*}
\Du\psi^{\l}_{\s} & = \delta^{\l}(u)\psi^{\l}_{\s}\,, \\
y_i\psi^{\l}_{\s} & = \qi \psi^{\l}_{\s}.
\end{align*}
Since the joint spectrum of the operators $\Du$ and $y_i \; \iN$ is
multiplicity-free, we must have:
\begin{equation*}
\psi^{\l}_{\s} \propto \ph\,.
\end{equation*}
This contradicts the observation that the maximal monomial of $\psi^{\l}_{\s}$
is smaller than any of the monomials $\mon \quad (\s \in \S)$.

Thus $s=0$  therefore $ \psi^{\l}_{\s} = \ph $ and consequently  $ \ph =
{\lim}_{p\rightarrow 1}\Ph\quad (\l \in \Lambda_N, \s \in \S )$.

\subsection{Action of the Hecke Algebra in the eigenspaces ${\cal E}^{\l}\;(\l
\in \Lambda_N)$ of the
operator $\Du$ .}

{\bf 1.} According to (3.3.4) the Hecke Algebra $\hecke$ generated by $ \gi \;
(i=1,\dots,N-1) $ acts in each eigenspace ${\cal E}^{\l}\;(\l \in \Lambda_N)$
of the operator $\Du$. To compute this action explicitely in the basis $\{ \ph
\}_{\s \in \S}$ we can either repeat almost word-by-word the derivation
described in {\bf 1.5 } or take the limit $p \rightarrow 1$ in the result of
Proposition 3. Either way we arrive at the following proposition:
\begin{prop} Let $\s \in \S \quad (\l\in \Lambda_N)$, then  $\hecke$ generated
by $ \gi \; (i=1,\dots,N-1) $ acts in  ${\cal E}^{\l} = \cplx \{ \ph \}_{\s \in
\S}$ as follows:
\begin{equation}
\gi\ph = \frac{\d \qii}{\qii-\qi}\ph +
\begin{cases}
& q\frac{(q^{-1}\qii-q\qi)(q\qii-q^{-1}\qi)}{(\qii-\qi)(\qii-\qi)}\phh  \\
& \text{{\em when} $ \l_{\s_i} > \l_{\s_{i+1}} \Rightarrow (i,i+1)\s \in \S $}
,  \\
& 0 \quad \text{{\em when} $\l_{\s_i} = \l_{\s_{i+1}} \Leftrightarrow (i,i+1)\s
\not\in \S $},  \\
& q^{-1}\phh  \\
& \text{{\em when} $ \l_{\s_i} < \l_{\s_{i+1}} \Rightarrow (i,i+1)\s \in \S $}.
\end{cases}
\end{equation}
\end{prop}

\subsection{``Motifs'' and associated partitions}

{\bf 1.} Following [HHTBP,BPS]  introduce the definition:
\begin{df}
Call a sequence of $M$ integers $(m_1,m_2,\dots,m_M)$ a {\em motif} iff:
\begin{align}
1 \leq m_1 < m_2 < \ldots < m_M \leq N-1 \, , \tag{i} \\
m_{i+1} \geq m_i + 2 \quad (i=1,\dots,M-1). \tag{ii}
\end{align}
\end{df}
With any motif $(m_1,m_2,\dots,m_M)$ associate a partition $\mu$:
\begin{equation*}
\mu = (M,\dots,\underset{m_1}{M},M-1,\dots,\underset{m_2}{M-1}, \; \ldots \;
,\underset{m_{M-1}+1}{1},\dots,\underset{m_M}{1},0,\dots,\underset{N}{0}) .
\end{equation*}
The subscripts in the last equation indicate positions of numbers in the
partition.

In what follows we shall identify  motifs with the partitions they define. We
shall indiscriminately use the notation  $(m_1,m_2,\dots,m_M)$ for both a motif
and the corresponding partition. Let $ \Mn $ be the set of all motifs for a
fixed $N$. We use the same notation for the corresponding subset of all
partitions.

\mbox{}

\noindent {\bf 2.} Let $\m$ be a partition from the set $ \Mn $. We subdivide
the set $\Sm$ (cf. {\bf 1.4 }) into disjoint subsets:
\begin{df}
For any subset $\seti \subset \setM \quad \LM  $  define  $ \Smi \subset \Sm
\subset S_N $ as follows:
\begin{eqnarray*}
\lefteqn{S_{N,(\emptyset)}^{\m}   := \{ {\mathrm {id}} \} , } \\
\lefteqn{\text{{\em for } $ 1 \leq L \leq M $ } \qquad  \Smi  := } \\
& & \left\{ \s \in \Sm \quad \begin{array}{|c c}  p^{\s}_{m_{i_k}} >
p^{\s}_{m_{i_k + 1}}  &  \forall\;  1\leq k \leq L \\
 p^{\s}_{m_j} <  p^{\s}_{m_j + 1 }  & \forall\;  j \in \setM\setminus\seti
\end{array} \right\}
\end{eqnarray*}
\end{df}
Recall ({\bf 1.4}) that for $ \s \in S_N $ we define:
\begin{equation*}
 \{ \s_1,\dots,\s_N \} := \s.\setN \; ,\quad i = \s_{p^{\s}_i} \quad \iN .
\end{equation*}

\noindent {\bf Example} Let $N=4\;,\; M=2$ and the motif is: $ (m_1 , m_2) = (1
, 3)$. The corresponding partition is: $ (2,1,1,0) $. In this case the set $
S_N^{(m_1,m_2)} = S_4^{(1,3)} $ contains altogether twelve elements. This set
is subdivided into four subsets: $ S_{4,(\emptyset)}^{(1,3)}\;,\;
S_{4,(1)}^{(1,3)}\;,\; S_{4,(3)}^{(1,3)}\;,\; S_{4,(1,3)}^{(1,3)} $ :
\begin{align*}
S_{4,(\emptyset)}^{(1,3)} & = \{ \{ 1234 \} \} \,; \\
S_{4,(1)}^{(1,3)} & = \{ \{ 2134\} ,\{ 2314 \} ,\{ 2341 \} \} \, ; \\
S_{4,(3)}^{(1,3)} & = \{ \{ 1243 \} ,\{ 1423 \} ,\{ 4123 \} \} \, ;  \\
S_{4,(1,3)}^{(1,3)} & = \{ \{ 2143\} ,\{ 2413 \}  ,\{ 4213 \} ,\{ 2431 \} ,\{
4231 \} \} .
\end{align*}

\mbox{}

\noindent {\bf 3.} Let us describe several properties of the sets $\Smi \quad (
\m \in \Mn \;, \seti \subset \setM \quad \LM )$. Throughout this paragraph we
fix such a set $\Smi $. Let $\mu$ be the partition that corresponds to $\m$.
\begin{lemma}
Let $\s \in \Smi $.

then either:
\begin{align*}
\exists \: i \in \{1,2,\dots,N-1 \} \, , & \;  \s\prime  \in \Smi \quad
\text{{\em s.t.}}   \tag{i} \\
& \s = (i,i+1)\s\prime \: ,\; \mu_{\s_i} < \mu_{\s_{i+1}} . \\
& \text{{\em and therefore} $ \s\prime > \s \; ( \Leftrightarrow \mu_{\s\prime}
> \mu_{\s})$};
\end{align*}
or
\begin{align*}
\s & = \s[0]:= (m_{i_1},m_{i_1}+1)(m_{i_2},m_{i_2}+1)\ldots (m_{i_L},m_{i_L}+1)
 . \tag{ii}
\end{align*}
\end{lemma}
\begin{pf}

\noindent Let $ \s = \{\s_1,\s_2,\dots,\s_N\} $. Examine the pairs $ \s_i ,
\s_{i+1} \; (i =1,\dots,N-1) $ step by step starting with $i=1$ and increasing
$i$ by $1$ at every next step.

\noindent At each step $i$ one has the two possibilities:
\begin{equation*}
{\mathrm  I.}\quad  \mu_{\s_i} \geq \mu_{\s_{i+1}} \qquad \qquad {\mathrm  II.}
 \quad \mu_{\s_i} < \mu_{\s_{i+1}}
\end{equation*}

\noindent If  {\em I.}, go to the next step. If at each step $ (i
=1,\dots,N-1)$ holds {\em I.}, then $\s = {\mathrm id}$ and therefore $ L=0 \,
, \, \s[0] = {\mathrm id} $. The poof is finished.

\noindent If  {\em II.}, then one has the further two possibilities:
\begin{equation*}
1. \; (i,i+1)\s \not\in \Smi  \qquad \qquad  2.  \;  (i,i+1)\s \in \Smi .
\end{equation*}

\noindent If  {\em 1.} go to the next step. If for all cases where {\em II.}
holds we have {\em 1.}, then $ \s = \s[0]$. The poof is finished.

\noindent If  {\em 2.} denote $\s\prime := (i,i+1)\s $. Since $  \mu_{\s_i} <
\mu_{\s_{i+1}} $, we have  $\mu_{\s\prime} > \mu_{\s} \Leftrightarrow
\s\prime > \s .$ The poof is finished.
\end{pf}

\noindent The element $\s[0]: = (m_{i_1},m_{i_1}+1)(m_{i_2},m_{i_2}+1)\ldots
(m_{i_L},m_{i_L}+1) \\   \in \Smi $ is the maximal element in the set $\Smi$
(Cf. {\bf 1.4.2} for the definition of the ordering of elements of $\Sm$).

\noindent We summarize the properties of the subset $\Smi $ in the following
proposition:
\begin{prop}
Let $\seti \subset \setM \;,\; \LM $.

\noindent Then:
\begin{equation*}
\Smi  = S_{N,(m_{i_1},\dots,m_{i_L})}^{(m_{i_1},\dots,m_{i_L})}\; ,\tag{i}
\end{equation*}
\begin{gather*}
\forall \; \s  \in \Smi \quad  \exists \;  \{ j_1,j_2,\dots,j_r\} \subset \{
1,\dots,N-1\} \quad ( r \geq 0 )  \; \text{{\em s.t.}} \tag{ii}
\end{gather*}
the elements $ \s[r] , \s[r-1],\dots , \s[1] $ defined by
\begin{align*}
\s[0] & := (m_{i_1},m_{i_1}+1)(m_{i_2},m_{i_2}+1)\dots (m_{i_L},m_{i_L}+1),  \\
\s[k] & := (j_k,j_k+1)\s[k-1] \quad ( k=1,2,\dots,r )
\end{align*}
belong to the set $ \Smi $, \\
satisfy
\begin{equation*}
       \s[k] < \s[k-1] \qquad ( k=1,2,\dots,r ),
\end{equation*}
and $ \s[r] = \s $. \\

\mbox{}

\noindent {\em (iii)} \\
If there exists $i \in \{1,2,\dots,N\}$ such that
\begin{align*}
  \s \; , \; (i,i+1)\s \; & \in \; \Smi  ,   \\
\text{then} &  \quad |\s_{i+1} - \s_i | \geq 2 .
\end{align*}
If there exists $i \in \{1,2,\dots,N\}$ such that
\begin{align*}
 \s \;\in \; \Smi , & \; (i,i+1)\s \;  \in \;
S_{N,(m_{i_2},\dots,m_{i_L})}^{\m}  ,   \\
\text{then} &  \quad  \s_i  \s_{i+1} + 1 .
\end{align*}
\end{prop}
\begin{pf}

\noindent {\em (i)} is a direct consequence of the {\em Definitions 1 ({\bf
1.4}) and 3}.

\noindent {\em (ii)} follows from Lemma 3 and the observation that $\s[0] =
(m_{i_1},m_{i_1}+1)(m_{i_2},m_{i_2}+1)\ldots (m_{i_L},m_{i_L}+1) $ is the
maximal element in $\Smi $.

\noindent {\em (iii)} For $\s \in \Smi $ assume that $ \s_{i+1} = \s_i +1 $.
One has the two possibilities:
\begin{align*}
1. & \quad i = p^{\s}_j \; , \; i+1 = p^{\s}_{j+1} \quad \text{where}\;
m_{s-1}+1 \leq j,j+1\leq m_s \\
 & \text{for some}\;  s \in \setM \; (m_0 := 0).  \\
2. & \quad i = p^{\s}_{m_s}\; , \; i+1 = p^{\s}_{m_s+1}  \\
& \text{for some} \; s \in \setM .
\end{align*}
In the case  {\em 1.} $ (i,i+1)\s \not\in \Sm .$  \\
In the case  {\em 2.} $ (i,i+1)\s \in S_{N,(m_s,m_{i_1},\dots,m_{i_L})}^{\m} .
$

\noindent Assume that $ \s_i = \s_{i+1} + 1 .$

\noindent In this case $ \exists \; s \in \seti $ {\em s.t. }
\begin{equation*}
\s_i = m_s + 1 \; , \; \s_{i+1} = m_s \quad \Leftrightarrow \quad i =
p^{\s}_{m_s+1} \; , \;i +1 = p^{\s}_{m_s}.
\end{equation*}
Since by definition of $\Smi$ :
\begin{gather*}
i < p_{m_s+2}  < \ldots < p_{m_{s+1}}   \, , \\
p_{m_{s-1}}  < \ldots < p_{m_s-2} < p_{m_s-1}< i+1  ;
\end{gather*}
we have:
\begin{equation*}
(i,i+1)\s \in S_{N,(m_{i_1},\dots,m_{i_L})\setminus m_s}^{\m}.
\end{equation*}
The first statement in {\em (iii)} is proven.

\noindent The second statement in {\em (iii)} is proven in a similar way.
\end{pf}

\subsection{A property of the eigenvectors $\ph$ for $\l \in \Mn$}

{\bf 1.} In this section we shall derive a certain proprerty of the eigenspaces
${\cal E}^{\mu}\quad (\mu \in \Mn)$ of the operators $\Du , y_i \iN. $
First of all, we notice that the eigenvalue $ \delta^{\mu}_1(u) $ of $ \Du $
associated with ${\cal E}^{\mu}$ can be represented in the additive
``particle'' form:
\begin{equation}
\delta^{\mu}_1(u) := \delta^{\m}_1(u) = \sum_{k=1}^M \delta^{(m_k)}_1(u),
\end{equation}
where $ \delta^{(m)}_1(u) \quad m \in \{1,\dots,N-1\}$ is the  one-particle
eigenvalue:
\begin{equation}
\delta^{(m)}_1(u) := u\sum_{i=1}^m \prod\begin{Sb} 1\leq j\leq N \\ j\neq i
\end{Sb} (1 + u q^{l_j})q^{l_i} .
\end{equation}
(Cf. 2.2.26 for the definition of $ l_i $ ).

\noindent {\bf 2.} Let $ \mu = \m \in \Mn$ then the conjugate partition
$\mu\prime$ is: $\mu\prime = (m_M,m_{M-1},\dots,m_1)$.
Due to (2.2.9,.10) we have:
\begin{equation}
\Du e_{m_1}e_{m_2}\dots e_{m_M} = \left(\sum_{k=1}^M
\delta^{(m_k)}_1(u)\right)e_{m_1}e_{m_2}\dots e_{m_M}.
\end{equation}
While for any symmetric polynomial $S$ one has:
\begin{equation}
y_i S = q^{l_i} S \quad \iN .
\end{equation}

\mbox{}

\noindent Now we have the following proposition:
\begin{prop}
Let $ \m \in \Mn\:,\; \seti \subset \setM \;\\ \LM, $ and $
\{j_1,\dots,j_{M-L}\} = \setM\setminus\seti. $

\noindent Then:
\begin{equation}
\varphi^{\m}_{\s} \; (\s \in \Smi)\; = \varphi^{(m_{i_1},\dots,m_{i_L})}_{\s}
e_{m_{j_1}}e_{m_{j_2}}\dots e_{m_{j_{M-L}}}.
\end{equation}
\end{prop}

\begin{pf}
Compute the action of the operators $\Du , y_i \; \iN. $ on the polynomial $
\varphi^{(m_{i_1},\dots,m_{i_L})}_{\s} e_{m_{j_1}}e_{m_{j_2}}\dots
e_{m_{j_{M-L}}}.$
\begin{align}
y_i\varphi^{(m_{i_1},\dots,m_{i_L})}_{\s}S &  =
q^{l_{\s_i}}\varphi^{(m_{i_1},\dots,m_{i_L})}_{\s}S \quad \iN. \tag{a}
\end{align}
For any symmetric polynomial $ S $.
\begin{multline}
\Du\varphi^{(m_{i_1},\dots,m_{i_L})}_{\s}e_{m_{j_1}}e_{m_{j_2}}\dots
e_{m_{j_{M-L}}} = \\
= [\Du\,,\,\varphi^{(m_{i_1},\dots,m_{i_L})}]e_{m_{j_1}}e_{m_{j_2}}\dots
e_{m_{j_{M-L}}} + \tag{b} \\
+ \varphi^{(m_{i_1},\dots,m_{i_L})}\Du e_{m_{j_1}}e_{m_{j_2}}\dots
e_{m_{j_{M-L}}}.
\end{multline}
Due to the special form of $\Du$ (Cf. {\em Proposition 4}) the commutator in
the last formula is a zero-order differential operator. Therefore if $ S $ is a
symmetric polynomial we have:
\begin{equation}
[\Du\,,\,\varphi^{(m_{i_1},\dots,m_{i_L})}]S = f(u)S,
\end{equation}
where $f(u)$ is a function of $ z_i \; \iN $ independent of $S$.
Furthermore:
\begin{equation*}
\Du 1 = 0,
\end{equation*}
and therefore:
\begin{align*}
[\Du\,,\,\varphi^{(m_{i_1},\dots,m_{i_L})}]1 &  =\Du
\varphi^{(m_{i_1},\dots,m_{i_L})} = \\
 & = \sum_{1\leq k\leq L}^M
\delta^{(m_{i_k})}_1(u)\varphi^{(m_{i_1},\dots,m_{i_L})}.
\end{align*}
Taking {\em (b),(4.4.7,.11)} into account we get:
\begin{multline*}
\Du\varphi^{(m_{i_1},\dots ,m_{i_L})}_{\s}e_{m_{j_1}}e_{m_{j_2}}\dots
e_{m_{j_{M-L}}} =  \\
= \left(\sum_{1\leq k\leq L} \delta^{(m_{i_k})} + \sum_{1\leq s\leq M - L}
\delta^{(m_{j_s})} \right) \varphi^{(m_{i_1},\dots
,m_{i_L})}_{\s}e_{m_{j_1}}e_{m_{j_2}}\dots e_{m_{j_{M-L}}}=  \\
= \sum_{1\leq i\leq M} \delta^{(m_i)} \varphi^{(m_{i_1},\dots
,m_{i_L})}_{\s}e_{m_{j_1}}e_{m_{j_2}}\dots e_{m_{j_{M-L}}}.
\end{multline*}

\noindent Since the joint spectrum of $ \Du \,,\,y_i \; \iN $ is
multiplicity-free, we conclude from {\em (a)} and the last equation that:
\begin{equation}
\varphi^{(m_{i_1},\dots,m_{i_L})}_{\s}e_{m_{j_1}}e_{m_{j_2}}\dots
e_{m_{j_{M-L}}} = {\mathrm {const}}\, \varphi^{\m}_{\s} \quad ( \s \in \Smi ) .
\end{equation}
By comparison of the maximal monomials in the both sides of the equation {\em
const} $ =1$.
\end{pf}

\mbox{}

\noindent Since $ e_r ( z_1=\omega^1,\dots,z_N = \omega^N ) = 0 \quad (1\leq r
\leq N-1)$
we have the following corollary to  Proposition 9:
\begin{cor}
Let $ \m \in \Mn\:,\; M > 0 \:,\;  \s \in \Sm  $.

\noindent Then:
\begin{align*}
\varphi^{\m}_{\s}( z_1=\omega^1,\dots,z_N = \omega^N )& = 0   \\
\text{{\em unless}} & \quad \s \in S^{\m}_{N,\m}.
\end{align*}
\end{cor}

\subsection{Limit $q\rightarrow 0$ of the eigenfunctions of $\Du , y_i\quad
(i=1,\dots,N)$}

{\bf 1.} The aim of this subsection is to compute some of the eigenfunctions
$\ph\quad (\l\in \Mn,\s \in \S)$ in the limit $q\rightarrow 0$ {\em under
assumption} that this limit is well-defined. We do not have a proof of the last
statement. In particular examples where $N$ is small this statement holds.

\mbox{}

\noindent Introduce operators:
\begin{equation}
\gamma_{i,j} := qg_{i,j}|_{q=0} = \frac{z_i}{z_i -z_j}(K_{i,j}- 1) \quad (i\neq
j\in \{1,\dots,N\} ).
\end{equation}
\begin{lemma}
Let $\m \in \Mn$, and $\psi\in\Cz$ satisfies the equations:
\begin{align}
(\gamma_{m_i,m_i+1} + 1 )\psi & = 0 \quad ( i=1,\dots,M), \\
\gamma_{n,n+1}\psi & = 0 \quad ( n\in
\{1,\dots,N-1\}\setminus\{m_1,\dots,m_M\}).
\end{align}
Then
\begin{equation}
\psi = (z_1\dots z_{m_1})^M(z_{m_1+1}\dots
z_{m_2})^{M-1}\ldots(z_{m_{M-1}+1}\dots z_{m_M}) S(z_1,\dots,z_N) ,
\end{equation}
where $S$ is a symmetric polynomial.
\end{lemma}
\begin{pf}
Let $N=2$ , $\psi = \psi(z_1,z_2)$.

The equation $ (\gamma_{1,2}+ 1)\psi = 0 $ implies that
\begin{equation}
\frac{z_1\psi(z_2,z_1) - z_2\psi(z_1,z_2)}{z_1 - z_2} = 0.
\end{equation}
This leads to $ \psi(z_1,z_2) = z_1 S(z_1,z_2)$, where $S$ is a symmetric
polynomial.

The equation
\begin{equation}
 \gamma_{1,2}\psi = \frac{z_1(\psi(z_2,z_1) - \psi_(z_1,z_2))}{z_1 - z_2} = 0 ,
\end{equation}
yields $ \psi(z_1,z_2) = S(z_1,z_2) $ where $ S$ is a symmetric polynomial.

The case of arbitrary $N$ reduces to the case $ N=2$ by consideration of
consequtive pairs of coordinates.
\end{pf}

\mbox{}

\noindent {\bf 2.} We {\em conjecture} that the limit $q \rightarrow 0$ of
$\ph$ is well-defined:
\begin{equation}
\ph|_{q=0}:= \varphi_{\s}^{\l,0} = z^{\l_{\s}} + \text{``s.m.''} \qquad (\l \in
\Mn ,\; \s\in \S).
\end{equation}
In what follows we assume that the statement of the conjecture is valid.

Fix $\m \in \Mn $. Our purpose is to find $ \varphi_{\s[0]}^{\m,0}$ for $
\s[0]:= (m_1,m_1+1)\dots(m_M,m_M+1).$ Take the limit $q\rightarrow 0$ in the
eq. (4.4.4) of Proposition 9. This yields
\begin{align}
(\gamma_{m_i,m_i+1} + 1) \varphi_{\s[0]}^{\m,0} & =
\varphi_{(m_i,m_i+1)\s[0]}^{\m,0}\quad (i = 1,\dots,M),\\ \gamma_{n,n+1}
\varphi_{\s[0]}^{\m,0} & = 0 \quad (n \in
\{1,\dots,N-1\}\setminus\{m_1,\dots,m_M\}).
\end{align}
According to Proposition 12, we have
\begin{multline}
\varphi_{(m_i,m_i+1)\s[0]}^{\m,0} =
\varphi_{(m_1,m_1+1)\dots\widehat{(m_i,m_i+1)}\dots (m_M,m_M+1)}^{\m,0} = \\
\varphi_{(m_1,m_1+1)\dots\widehat{(m_i,m_i+1)}\dots
(m_M,m_M+1)}^{(m_1,\dots,\hat{m_i},\dots,m_M ),0} e_{m_i} \quad (i =1,\dots,M).
\end{multline}
Where we put a hat over terms that are omitted.

The pair of equations (4.4.20,.21) provides a set of recurrent relations for
the eigenfunctions $\varphi_{\s[0]}^{\m,0}$. Notice that when $M=0$ we have $
\varphi_{\s[0]}^{(\emptyset),0} = 1$. Taking into account Lemma 4 we write the
general solution of these recurrent relations:
\begin{align}
\varphi_{\s[0]}^{(\emptyset),0} &  = 1, \\
\varphi_{\s[0]}^{\m,0} & = (e_{m_1} - z_1\dots z_{m_1})(e_{m_2} - z_1\dots
z_{m_2})\ldots (e_{m_M} - z_1\dots z_{m_M})+  \notag\\ & + (z_1\dots
z_{m_1})^M(z_{m_1+1}\dots z_{m_2})^{M-1}\ldots(z_{m_{M-1}+1}\dots
z_{m_M})\times \notag \\ & \times  S^{\m}(z_1,\dots,z_N) \quad ( M \geq 1).
\notag
\end{align}
Where $  S^{\m}(z_1,\dots,z_N) $ is an arbitrary symmetric polynomial.

Observe now that the total degree of the homogeneous polynomial $
\varphi_{\s[0]}^{\m,0} $ is equal to $ m_1 + \dots + m_M  \quad (\m \in \Mn) $.
Therefore we must have $   S^{\m}(z_1,\dots,z_N) = {\mathrm {const}} $. Observe
further that
\begin{multline}
  (z_1\dots z_{m_1})^M(z_{m_1+1}\dots z_{m_2})^{M-1}\ldots(z_{m_{M-1}+1}\dots
z_{m_M}) > \\
 > {\mathrm {max}}( \varphi_{\s[0]}^{\m,0} ) \quad ( \s[0]:=
(m_1,m_1+1)\dots(m_M,m_M+1) ).
\end{multline}
Where $ {\mathrm {max}}( \varphi_{\s[0]}^{\m,0} ) $ is the maximal monomial of
$ \varphi_{\s[0]}^{\m,0}. $
Therefore we must have: $ S^{\m}(z_1,\dots,z_N)  = 0 $.

Thus for $ \m \in \Mn $ :
\begin{multline}
\varphi_{\s[0]}^{(\emptyset),0}   = 1, \\
\varphi_{\s[0]}^{\m,0}  = (e_{m_1} - z_1\dots z_{m_1})(e_{m_2} - z_1\dots
z_{m_2})\ldots (e_{m_M} - z_1\dots z_{m_M})\notag  \\  ( M \geq 1 ) \notag
\end{multline}

Notice that
\begin{multline}
\varphi_{\s[0]}^{\m,0}(z_1=\omega^1,\dots,z_N=\omega^N ) = (-1)^M
\omega^{\frac{1}{2}\sum_{i=1}^M m_i(m_i+1)} , \\ \quad ( \m \in \Mn , \; \s[0]=
(m_1,m_1+1)\dots(m_M,m_M+1)).
\end{multline}

\section{Hecke-invariant (``bosonic'') subspaces of ${\cal E}^{\l}\otimes H$
for ${\l} \in \Mn $}

\subsection{Preliminaries}

{\bf 1.} Let ${\cal E}^{\l} \subset \Cz$ be the eigenspace of the operators $
\Du\,,\,y_i\quad\iN$ parametrized by a partition $\l$ and $ H :=
(\cplx^2)^{\otimes N}$. The bosonic subspace $\B^{\l}$ of ${\cal E}^{\l}\otimes
H $ is defined as follows:
\begin{equation}
B^{\l}:= \{ b \in {\cal E}^{\l}\otimes H | (\gi -\ti)b = 0\quad \iNN \}.
\end{equation}
Since $ {\cal P}:=\Cz\otimes H = \oplus_{\l}({\cal E}^{\l}\otimes H)$ and $ \gi
:{\cal E}^{\l}\mapsto {\cal E}^{\l}\quad\iNN ;$ we have: $ \B
=\oplus_{\l}\B^{\l}. $ (Cf. (1.1.36), (3.3.7) for the definition of $\B$).

\noindent {\bf 2.} Any vector $\psi$ from ${\cal E}^{\l}\otimes H$ is
represented as follows:
\begin{equation}
\psi = \sum_{\s\in\S}\ph \x ,
\end{equation}
where $\x \quad (\s \in \S) \in H $.

\mbox{}

\noindent The condition $ (\gi -\ti)\psi = 0\quad\iNN  $ gives a set of linear
equations which must be satisfied by the vectors   $\x \quad (\s \in \S) \in H
.$ In order to derive these equations we can apply the result of Proposition 9
to find out the action of $\gi\quad\iNN  $  on $\psi$, and then use the
linear-independence of the polynomials $ \ph\quad (\s\in\S)$. In this way we
arrive at the following proposition:
\begin{prop}
A vector $ \psi \in {\cal E}^{\l}\otimes H  ;\; \psi =  \sum_{\s\in\S}\ph \x $
belongs to $ \B^{\l} $ iff $\x \quad (\s \in \S) \in H $  satisfy the following
set of equations $ \iNN $:
\begin{gather*}
q\frac{(q^{-1}\qii - q\qi)(q^{-1}\qi -q\qii)}{\qi - \qii}\xx  =\tag{a} \\ =
\left( (\qii-\qi)\ti - \d\qii\right)\x \\
 \text{{\em when}}\quad \l_{\s_{i+1}} > \l_{\s_i }\;, \\
(\ti - q ) \x  = 0 \tag{b}  \\
 \text{{\em when}}\quad \l_{\s_{i+1}} = \l_{\s_i } \Leftrightarrow (i,i+1)\s
\not\in \S \Rightarrow \s_{i+1} = \s_i+1 \; , \\
\xx = \frac{ (\qii -\qi)\ti - \d\qii}{q^{-1}(\qii - \qi)}\x  \tag{c}\\
 \text{{\em when}}\quad \l_{\s_{i+1}} < \l_{\s_i } \; .
\end{gather*}
\end{prop}

\subsection{ Spaces $\B^{\mu}$ for $ \mu \in \Mn $ }

{\bf 1.} Let us fix a partition $\mu \in \Mn$ parametrized by a motif $ \m $.
We use the same  notation $\m$ for both the motf and the partition.

\noindent For $\mu \in \Mn$ let us further analyse the equations (a)-(c)
obtained in the Proposition 10. In section {\bf 4.3} (Definition 3) we
introduced the decomposition of the set $ \Sm $ into disjoint subsets:
\begin{equation}
\Sm = \bigsqcup_{\seti \in \setM} \Smi \qquad  \LM
\end{equation}
This decomposition is reflected in the equations (a)-(c) of the Proposition 10.

\noindent {\bf 2.} Let in these equations $ \s\,,\, i $ be such that $ \s \,
,\, (i,i+1)\s \in \Smi $ for some $ \seti \in \setM $. According to Proposition
8,(iii) $ |\s_i - \s_{i+1}| \geq 2.$

Let $ \s_i - \s_{i+1} \geq 2 $ and consequently $ \mu_{\s_i} < \mu_{\s_{i+1}}
$. Since $ \s_i - \s_{i+1} \geq 2 $ the coefficient in front of $ \xx $ in
Pr.10(a) is not equal to zero. Therefore we have:
\begin{multline}
\xx = \frac{(\qi-\qii)((\qii-\qi)\ti
-\d\qii)}{q(q^{-1}\qii-q\qi)(q^{-1}\qi-q\qii)}\x   \\
( \s , (i,i+1)\s \in \Smi\, ,\, \s_i - \s_{i+1} \geq 2 ).
\end{multline}

\mbox{}

\noindent Let $\s_i - \s_{i+1} \leq -2$ and consequently  $ \mu_{\s_i} >
\mu_{\s_{i+1}} $. Pr.10(c) gives:
\begin{multline}
\xx = \frac{((\qii-\qi)\ti -\d\qii)}{q^{-1}(\qii-\qi)}\x   \\
( \s , (i,i+1)\s \in \Smi\, ,\, \s_i - \s_{i+1} \leq  -2 ).
\end{multline}

\mbox{}

\noindent Introduce a pair of $U$-intertwiners:
\begin{equation}
Y^{\pm}(z) : = \varrho^{\pm}(z)\frac{zt - t^{-1}}{q^{-1}z - q} \; \in
End(V\otimes V),
\end{equation}
where
\begin{equation}
\varrho^+(z) := \frac{z - 1}{q^2z - 1 }\; ,  \qquad \varrho^-(z) := \frac{z -
q^2}{z - 1 }.
\end{equation}
The eq. (5.5.4,.5) can be reformulated as follows:
Let $ \s , (i,i+1)\s \in \Smi \\ \quad (\seti \in \setM \quad \LM ) $ then:
\begin{equation}
\xx =
\begin{cases}
Y^+_{i,i+1}(q^{l_{\s_i}-l_{\s_{i+1}}}) \x & \text{when $ \s_i - \s_{i+1} \geq
2 \Rightarrow \mu_{\s_i} < \mu_{\s_{i+1}}$ }, \\
Y^-_{i,i+1}(q^{l_{\s_i}-l_{\s_{i+1}}}) \x & \text{when $ \s_i - \s_{i+1} \leq
-2 \Rightarrow \mu_{\s_i} > \mu_{\s_{i+1}}$}.
\end{cases}
\end{equation}
Notice that when  $ \s , (i,i+1)\s \in \Smi $ the intertwiners
$Y^{\pm}_{i,i+1}(q^{l_{\s_i}-l_{\s_{i+1}}})$ are invertible.

\mbox{}

\noindent {\bf 3.} Now consider the situation when $ \s$ and $(i,i+1)\s$ in
Proposition 10 belong to different subsets $\Smi$. According to Proposition 8
this situation takes place when $|\s_i - \s_{i+1} | = 1.$

\noindent Let $ \s_{i+1}= \s_i + 1$ and consequently $ \mu_{\s_i} >
\mu_{\s_{i+1}}.$ In this case $ \exists \; s \in \setM $ s.t. $ \s_i = m_s \, ,
\: \s_{i+1} = m_s + 1 $ or , equivalently , $ i = p^{\s}_{m_s} \,,\:   i+1 =
p^{\s}_{m_s+1}$; and if $ \s \in \Smi $ then $ s \not\in \seti .$ Since $
((i,i+1)\s)_i = m_s+1\;,\; ((i,i+1)\s)_{i+1} = m_s ;$ we have $  (i,i+1)\s \in
S_{N,(m_s,m_{i_1},\dots,m_{i_L})}^{\m} $.

\noindent Substituting $ \s_{i+1}= \s_i + 1$ into Pr.10(c) we find:
\begin{multline}
\xx = q(\ti - q)\x = -(q^2 + 1)\Pi_{i,i+1}^-(q)\x    \\
 ( \s \in \Smi \;,\; (i,i+1)\s \in  S_{N,(m_s,m_{i_1},\dots,m_{i_L})}^{\m}\;,\;
\\  \s_i = m_s , \s_{i+1} = m_s + 1 ( s \in \setM\setminus\seti ) ).
\end{multline}
(Cf. (1.1.12,.13) for the definition of projectors $\Pi_{i,i+1}^{\pm}(q)$).

\mbox{}

\noindent Let $ \s_i= \s_{i+1} + 1$ and consequently $ \mu_{\s_i} <
\mu_{\s_{i+1}}.$ In this case $ \exists \; s \in \setM $ s.t. $ \s_i = m_s+1 \,
, \: \s_{i+1} = m_s $ ; and since $ \s \in \Smi \: ; \; s \in \seti .$ On the
other hand $ (i,i+1)\s \in  S_{N,(m_{i_1},\dots,m_{i_L})\setminus m_s}^{\m}.$
\\  Here $(m_{i_1},\dots,m_{i_L})\setminus m_s$ signifies the motif obtained
from $( m_{i_1},\dots,m_{i_L})$ by removing $m_s$.

\noindent Substituting $ \s_i= \s_{i+1} + 1$ into Pr.10(a) we find:
\begin{multline}
(\ti + q^{-1})\x = 0 \quad  \Rightarrow \quad  \Pi_{i,i+1}^+(q)\x = 0  \\
( \s \in \Smi \;,\; (i,i+1)\s \in  S_{N,(m_{i_1},\dots,m_{i_L})\setminus
m_s}^{\m}\;,\; \\  \s_i = m_s+1 , \s_{i+1} = m_s ( s \in \seti \subset \setM )
).
\end{multline}

\mbox{}

\noindent {\bf 4.} Now we are in position to reformulate Propositon 10 in a
more suitable form in the case when $ \l \equiv \mu  \equiv \m \in \Mn $.
\begin{prop}
Let $ \psi \in {\cal E}^{\mu}\otimes H$, i.e.
\begin{equation}
\psi = \sum_{\s \in \Sm}\varphi^{\m}_{\s} \x \qquad ( \x \in H ).
\end{equation}
Then $ \psi \in \B^{\m} $  iff  $  \x  \quad (\s \in \Sm) $ satisfy the
following linear relations:
\begin{multline}
\forall \; \seti \subset \setM \quad \LM   \\
\chi_{(m_{i_1},m_{i_1}+1)\dots (m_{i_L},m_{i_L}+1)} =  \\
= -(q^2 + 1) \Pi_{m_{i_k},m_{i_k}+1}^-(q) \chi_{(m_{i_1},m_{i_1}+1)\dots
\widehat{ (m_{i_k},m_{i_k}+1)}\dots (m_{i_L},m_{i_L}+1)} \\
(k = 1,2,\dots,L).
\end{multline}
Where  $\widehat{\mbox{}}$  means that the corresponding factor is omitted from
the product.
\begin{multline}
\forall \; \seti \subset \setM \quad \LM   \\
\text{{\em and }} \; j \in \{1,\dots,N-1\} \; \text{{\em s.t.}}  \\
\{j,j+1\}\cap \{m_1,m_1+1,m_2,m_2+1,\dots,m_M,m_M+1\} = \emptyset \; ; \\
\Pi_{j,j+1}^-(q)\chi_{(m_{i_1},m_{i_1}+1)\dots (m_{i_L},m_{i_L}+1)} = 0.
\end{multline}
\begin{gather}
\forall \; \s \in \Smi \quad ( \seti \subset \setM \quad \LM  )   \\
\x = \BY(\s)\chi_{(m_{i_1},m_{i_1}+1)\dots (m_{i_L},m_{i_L}+1)}. \notag
\end{gather}
Where invertible $\BY(\s)\;\in\; End(H)$ is recursively defined as follows:
\begin{gather*}
\BY((m_{i_1},m_{i_1}+1)\dots (m_{i_L},m_{i_L}+1)) := {\mathrm {Id}} \, , \\
\text{{\em for}} \; (i,i+1)\s \, \in \, \Smi  \\
\BY((i,i+1)\s) =
\begin{cases}
Y^+_{i,i+1}(q^{l_{\s_i}-l_{\s_{i+1}}})\BY(\s) & \text{{\em if}}\; \s_i -
\s_{i+1} \geq \mbox{} 2\, , \\
Y^-_{i,i+1}(q^{l_{\s_i}-l_{\s_{i+1}}})\BY(\s) & \text{{\em if}}\; \s_i -
\s_{i+1} \leq -2\, .
\end{cases}
\end{gather*}
\end{prop}

\mbox{}

\noindent It is possible to give more explicit expression for the matrix $
\BY(\s) $ that appears in (5.5.14). In notation of Proposition 8 (ii) we have:
\begin{align}
\text{For} & \; \s \in \Smi \qquad ( \seti \subset \setM \; \LM )  \notag \\
\BY(\s) & =  Y^-_{j_r,j_r+1}(q^{l_{\s [r-1]_{j_r}}-l_{\s [r-1]_{j_r+1}}}) \dots
 Y^-_{j_1,j_1+1}(q^{l_{\s [0]_{j_1}}-l_{\s [0]_{j_1+1}}}).
\end{align}
Where:
\begin{align}
\s&[k]   := (j_k,j_k+1)\dots (j_1,j_1+1)\s[0] \in \Smi \; ( 1\leq k \leq r),
\notag \\
\s&[0]  := (m_{i_1},m_{i_1}+1)\dots (m_{i_L},m_{i_L}+1) , \notag  \\
\s&[r] = \s \notag \\
\s&[k-1]_{j_k}   - \s[k-1]_{j_k+1} \leq -2 \quad (1 \leq k \leq r ).
\end{align}

\mbox{}

\noindent {\bf 5.} Introduce two definitions.\\
For $ \m \in \Mn $ :
\begin{equation}
{\cal Z}^{\m}:= \left\{ v \in H \; \begin{array}{|c } \Pi_{j,j+1}^-(q)v = 0
\quad \text{for all}\quad j\in\{1,\dots,N-1\}\;\text{s.t.} \\
 \{j,j+1\}\cap\{m_1,m_1+1,\dots,m_M,m_M+1\}=\emptyset \end{array} \right\}
\end{equation}
More explicitely:
\begin{multline*}
{\cal Z}^{\m} =
S_q(\underset{1}{V}\otimes\dots\otimes\underset{m_1-1}{V})\otimes\underset{m_1}{V}\otimes\underset{m_1+1}{V}\otimes S_q(\underset{m_1+2}{V}\otimes\dots\otimes\underset{m_2-1}{V})\otimes\underset{m_2}{V}\otimes\underset{m_2+1}{V}\otimes\ \\ \ldots \otimes S_q(\underset{m_M+2}{V}\otimes\dots\otimes\underset{N}{V}) \; \subset \; H := \underset{1}{V}\otimes\underset{2}{V}\otimes\ldots\otimes\underset{N}{V}.
\end{multline*}
Where $S_q$ means $q$-symmetrization and the subscripts indicate positions of
the factors in the tensor product $ V^{\otimes N}.$ \\
\mbox{} \\
For any $ \seti \in \setM \quad \LM $ define the following projector:
\begin{equation}
\Pi^{-,\m}_{(m_{i_1},\dots,m_{i_L})}:=
\begin{cases}\Pi^-_{m_{i_1},m_{i_1}+1}(q)\dots\Pi^-_{m_{i_L},m_{i_L}+1}(q) &
\quad (1\leq L \leq M), \\
 {\mathrm I} &  (L =0) .\end{cases}
\end{equation}

\mbox{}

\noindent {\bf 6.} Proposition 11 yields the following expression for $
\B^{\mu}\: \equiv \:\B^{\m} $:
\begin{equation}
\B^{\m} = \BU^{\m}(z){\cal Z}^{\m}.
\end{equation}
Where
\begin{multline}
\BU^{\m}(z) := \\  \sum\begin{Sb} I \subset \setM \\ I:=\seti
\end{Sb}(-(q^2+1))^L\left\{\sum_{\s\in\Smi}\varphi_{\s}^{\m}(z)\BY(\s)\right\}\Pi^{-,\m}_{(m_{i_1},\dots,m_{i_L})} \; , \\
      \BU^{\m}(z) : H \mapsto \Cz\otimes H .
\end{multline}
In the last formula we explicitely indicated $z$-dependence of the polynomials
$\varphi_{\s}^{\m}$.

\mbox{}

\noindent {\bf 7.} The space $ \B^{\m}$ is a $U$-module where the action of $U$
is given by (2.2.19):
\begin{equation}
T_a^0(u) = L_a(u;\{y_i\}):= L_{a1}(uy_1)L_{a2}(uy_2)\dots L_{aN}(uy_N) \; \in
\; End(\Cz\otimes H).
\end{equation}
The space ${\cal Z}^{\m}$ is a (reducible,indecomposable) $U$-module as well,
with the $U$-action:
\begin{equation}
 L_a(u;\{q^{l_i}\}):= L_{a1}(uq^{l_1})L_{a2}(uq^{l_2})\dots L_{aN}(uq^{l_N}) \;
\in \; End(H).
\end{equation}
The operator $\BU^{\m}(z)$ in (5.5.19,.20) is an $U$-intertwiner of these two
modules. To see this let us consider the product  $T_a^0(u)\BU^{\m}(z)$.Since
$\varphi_{\s}^{\m}(z)$ are eigenvectors of $ y_i \; \iN $ (Cf. Proposition 6)
we get:
\begin{multline}
T_a^0(u)\BU^{\m}(z) = \\
\sum\begin{Sb} I \subset \setM \\ I:=\seti \end{Sb}(-(q^2+1))^L \; \times \\
\times \; \left\{\sum_{\s\in\Smi}\varphi_{\s}^{\m}(z)
L_a(u;\{\qi\})\BY(\s)\right\}\Pi^{-,\m}_{(m_{i_1},\dots,m_{i_L})}.
\end{multline}
It follows from the recursive definition of $\BY(\s)\quad (\s \in \Smi)$ given
in Proposition 11 that  $\BY(\s)$ is an $U$-intertwiner:
\begin{equation}
L_a(u;\{\qi\})\BY(\s) = \BY(\s)L_a(u;\{q^{l_{\s[0]_i}}\}).
\end{equation}
Where $ \s[0] = (m_{i_1},m_{i_1}+1)\dots (m_{i_L},m_{i_L}+1)$.\\
Furthermore the projector $\Pi^{-,\m}_{(m_{i_1},\dots,m_{i_L})}$ is an
interwiner as well (Cf. 1.1.28):
\begin{equation}
L_a(u;\{q^{l_{\s[0]_i}}\}\Pi^{-,\m}_{(m_{i_1},\dots,m_{i_L})} =
\Pi^{-,\m}_{(m_{i_1},\dots,m_{i_L})}L_a(u;\{q^{l_i}\}).
\end{equation}
Hence we obtain:
\begin{equation}
T_a^0(u)\BU^{\m}(z) = \BU^{\m}(z)L_a(u;\{q^{l_i}\}).
\end{equation}
Thus the intertwinig property of $ \BU^{\m}(z)$ is established.

\section{Spectrum and eigenspaces of the operators $\Xi^{(n)}(\omega) $ $
(n=1,\dots,N-1)$ forming the
hierarchy of $U$-invariant Spin Models.}

\subsection{Eigenspaces $H^{\mu}_{\bf{B}}(\omega)$ of the generating function
$\Xuo $}
{\bf 1.} Let $\mu \equiv \m \in \Mn .$ Define a subspace $H^{\mu}_{\B}(\omega)
\subset  H_{\B}(\omega) \subset H $ by applying the evaluation map
$Ev(\omega)$ (Cf. {\bf 3.1}) to the ``bosonic'' subspace $ \B^{\m} $ introduced
in the previous section:
\begin{equation}
H^{\m}_{\B}(\omega) := Ev(\omega)\B^{\m}\quad (\m \in\Mn).
\end{equation}

{}From (3.3.13,.14,.19) we obtain
\begin{multline}
\Xuo H^{\m}_{\B}(\omega) = \left(\sum_{i=1}^M (\delta^{(m_i)}_1(u) -
\theta(u)m_i)\right)H^{\m}_{\B}(\omega) \\   (\m \in\Mn).
\end{multline}
Where $ \delta^{(m_i)}_1(u) $ was defined in (4.4.6) and $\theta(u)$ was
defined in (2.2.6). The last equation says that $ H^{\m}_{\B}(\omega) $ is an
eigenspace of $\Xuo$ unless $ H^{\m}_{\B}(\omega) \equiv 0 $.

\mbox{}

\noindent {\bf 2.} Application of the evaluation map $ Ev(\omega) $ to
$\B^{\m}$ (5.5.19) yields the explicit expression for the space
$H^{\m}_{\B}(\omega)$:
\begin{equation}
H^{\m}_{\B}(\omega) = \BU^{\m}(\omega){\cal Z}^{\m}\quad (\m \in\Mn).
\end{equation}
Where
\begin{multline}
\BU^{\m}(\omega) := \BU^{\m}(z_1 =\omega^1,\dots,z_N = \omega^N) = \\
= (-(q^2+1))^M \sum_{\s\in S^{\m}_{N,\m}}\varphi_{\s}^{\m}(\omega)\BY(\s)
\Pi^{-,\m}_{\setM} \; , \\
      \BU^{\m}(\omega) : H \mapsto H .
\end{multline}
To get the expression for  $\BU^{\m}(\omega)$ we used the Corollary 1 to
Proposition 9.

\mbox{}

\noindent Let us introduce the notation
\begin{equation}
W^{\m} :=  \Pi^{-,\m}_{\setM}{\cal Z}^{\m}\quad (\m \in\Mn).
\end{equation}
The definition of ${\cal Z}^{\m}$ given in (5.5.17) leads to a more explicit
form of $W^{\m}$:
\begin{multline}
W^{\m} = \\ S_q(\underset{1}{V}\otimes\dots\otimes\underset{m_1-1}{V})\otimes
A_q(\underset{m_1}{V}\otimes\underset{m_1+1}{V})\otimes
S_q(\underset{m_1+2}{V}\otimes\dots\otimes\underset{m_2-1}{V})\otimes
A_q(\underset{m_2}{V}\otimes\underset{m_2+1}{V})\otimes\ \\ \ldots \otimes
S_q(\underset{m_M+2}{V}\otimes\dots\otimes\underset{N}{V}) \; \subset \; H :=
\underset{1}{V}\otimes\underset{2}{V}\otimes\ldots\otimes\underset{N}{V}.
\end{multline}
Where $A_q$ signifies $q$-antisymmetrization (1.1.15).

Observe that $W^{\m}\quad (\m \in\Mn)$ is an irreducible highest-wight
$U$-module with the $U$-action is given by
\begin{equation} L_a(u;\{q^{l_{\s[0]_i}}\}):=
L_{a1}(uq^{l_{\s[0]_1}})L_{a2}(uq^{l_{\s[0]_2}})\dots L_{aN}(uq^{l_{\s[0]_N}})
\; \in \; End(H).
\end{equation}
Where we used the notation
\begin{equation}
\s[0] := (m_1,m_1+1)(m_2,m_2+1)\ldots (m_M,m_M+1) \quad (\m \in\Mn).
\end{equation}
The Drinfeld polynomial of this module is
\begin{equation}
Q^{\m}(u) = \prod\begin{Sb}1\leq n \leq N \\ n \neq m_i,m_i+1 \end{Sb}(1 -
q^{-l_n}u).
\end{equation}

\mbox{}

\noindent According to (6.6.3) we have
\begin{multline}
H^{\m}_{\B}(\omega) = \Check{\BU}^{\m}(\omega)W^{\m}\quad (\m \in\Mn), \\
\Check{\BU}^{\m}(\omega):= (-(q^2+1))^M \sum_{\s\in
S^{\m}_{N,\m}}\varphi_{\s}^{\m}(\omega)\BY(\s).
\end{multline}

\mbox{}

\noindent {\bf 3.} The space $ H^{\m}_{\B}(\omega) \quad (\m \in\Mn)$ is a
$U$-module with the $U$-action given by $\Tuo$ defined in (3.3.19). Explicitely
(Cf. 5.5.23):
\begin{multline}
\Tuo H^{\m}_{\B}(\omega) = \Tuo\Check{\BU}^{\m}(\omega)W^{\m} = \\
=(-(q^2+1))^M \sum_{\s\in S^{\m}_{N,\m}}\varphi_{\s}^{\m}(\omega)
L_a(u;\{\qi\})\BY(\s), \\
 L_a(u;\{\qi\}):= L_{a1}(uq^{l_{\s_1}})L_{a2}(uq^{l_{\s_2}})\dots
L_{aN}(uq^{l_{\s_N}}).
\end{multline}
Applying (5.5.24) we find that $ \Check{\BU}^{\m}$ is an intertwiner of the
modules $H^{\m}_{\B}(\omega)$ and $ W^{\m}$:
\begin{multline}
\Tuo H^{\m}_{\B}(\omega) = \Tuo\Check{\BU}^{\m}(\omega)W^{\m} = \\
\Check{\BU}^{\m}(\omega)L_a(u;\{q^{l_{\s[0]_i}}\})W^{\m} \quad (\m \in\Mn).
\end{multline}

Since $ W^{\m}$ is irreducible so is  $H^{\m}_{\B}(\omega)$. The highest-weight
vector of  $ W^{\m}$ is
(Cf. 6.6.6)
\begin{multline}
\tilde{\Omega}^{\m} := \\ v^+\otimes\dots\otimes
v^+\otimes(\underset{m_1}{v^+}\otimes\underset{m_1+1}{v^-}-q v^-\otimes
v^+)\otimes \quad\ldots \quad\otimes v^+\otimes\dots\otimes v^+.
\end{multline}

\noindent If the vector $\Omega^{\m} := \Check{\BU}^{\m}\tilde{\Omega}^{\m} \in
H^{\m}_{\B}(\omega)$  is not zero, it is the highest-weight  vector,  and the
modules  $H^{\m}_{\B}(\omega)$ and  $ W^{\m}$ are isomorphic ,  with
$\Check{\BU}^{\m}$ defining the isomorphism explicitely.

\mbox{}

\noindent {\bf 4.} In order to show that $\Omega^{\m} \quad (\m \in\Mn) $ is
not zero we compute this vector at $q =0$.

\noindent Consider the matrix $\BY(\s)\quad (\s \in S_{N,\m}^{\m})$ that enters
the definition of $\Check{\BU}^{\m}$. According to (5.5.15,.16):
\begin{align}
\BY(\s) & = {\mathrm {Id}} \quad ( \s = \s[0]:= (m_1,m_1+1)\dots(m_M,m_M+1)),
\\
\BY(\s) & = \BY(\s)' Y^-_{i,i+1}(q^{l_{\s[0]_i}-l_{\s[0]_{i+1}}}) \quad ( \s
\neq \s[0]).
\end{align}
Where depending on $\s$ , $i$ takes one of the values in the set $
\{m_k-1,m_k+1\}_{k\in\setM}$. For any such $i$ we have $ \s[0]_i - \s[0]_{i+1}
= -2 .$ $\BY(\s)'$ is either identity or a product of intertwiners of the form
$ Y^-_{j,j+1}(q^{-2r}) \quad j\in (\{1,\dots,N-1\})$ where $ r \geq  2 . $

\noindent For $r \geq 2$ we find
\begin{equation}
Y^-_{j,j+1}(q^{-2r})|_{q = 0} = -\Pi_{j,j+1}^-(0) = -(|+-><+-|)_{j,j+1} \quad
(j=1,\dots,N-1).
\end{equation}
Thus
\begin{multline}
\lim_{q\rightarrow 0}\BY(\s) =  -\BY(\s)'|_{q=0}\Pi_{i,i+1}^-(0) \quad  (i\in
\{m_k-1,m_k+1\}_{k\in\setM}), \\ \quad (\s \in S_{N,\m}^{\m},\; \s\neq \s[0] ).
\end{multline}

The highest-weight vector $ \tilde{\Omega}^{\m}$ in the limit $ q\rightarrow 0
$ is
\begin{multline}
\tilde{\Omega}^{\m,q=0} := \\
v^+\otimes\dots\otimes\underset{m_1}{v^+}\otimes\underset{m_1+1}{v^-}\otimes
v^+\dots \otimes\underset{m_2}{v^+}\otimes\underset{m_2+1}{v^-}
\otimes\quad\ldots \quad\otimes v^+\otimes\dots\otimes v^+.
\end{multline}
Therefore taking into account (6.6.17) and (4.4.26) we arrive at the following
expression for $\Omega^{\m} \in H^{\m}_{\B}(\omega)$ at $q=0$:
\begin{multline}
\Omega^{\m,q=0} = (-1)^M \varphi_{\s[0]}^{\m}(\omega) \tilde{\Omega}^{\m,q=0} =
\\
=\omega^{\frac{1}{2}\sum_{i=1}^M m_i(m_i+1)} \tilde{\Omega}^{\m,q=0}.
\end{multline}
Since $\Omega^{\m,q=0}$ is not zero we can argue that same holds for any
generic value of $q$ (not a root of unity) and therefore $ H^{\m}_{\B}(\omega)
$ and $ W^{\m} $ are isomorphic $U$-modules for any generic $q$ and  $ \m \\
\in \Mn .$

\mbox{}

\noindent {\bf 5.} From consideration of the case $q=0$ we deduce that
\begin{equation}
H = \bigoplus_{\m \in \Mn} W^{\m}.
\end{equation}
Since the $U$-modules $ H^{\m}_{\B}(\omega)$ have different Drinfeld
polynomials (6.6.9) for different motifs $\m$ any two of these modules do not
intersect except at zero vector. Therefore we can take the direct sum of all
these modules. Since ${\mathrm{ dim}}H^{\m}_{\B}(\omega) = {\mathrm
{dim}}W^{\m} $, we conclude from (6.6.20) that
\begin{equation}
H = \bigoplus_{\m \in \Mn}  H^{\m}_{\B}(\omega) .
\end{equation}
Thus we have found the complete decomposition of the space of states into
eigenspaces of the operator $\Xuo$, as well as the $U$-representation content
of this decomposition.

\end{document}